\documentclass[lettersize,journal]{IEEEtran}
\usepackage{amsmath,amsfonts}
\usepackage{amsthm}

\usepackage{algorithmic}
\usepackage[linesnumbered,ruled,vlined]{algorithm2e}
\usepackage{array}
\usepackage{enumitem}
\usepackage{textcomp}
\usepackage{stfloats}
\usepackage{url}
\usepackage{tikz}
\usepackage{verbatim}
\usepackage{graphicx}
\usepackage{cite}
\usepackage{pifont}
\usepackage{caption}
\usepackage{booktabs}
\usepackage[table]{xcolor}
\usepackage[normalem]{ulem}
\usepackage{multirow}
\usepackage{orcidlink}
\usepackage{flushend}
\usepackage[most]{tcolorbox}
\usepackage[T1]{fontenc}
\usepackage{newtxtext,newtxmath}

\definecolor{lightblue}{RGB}{245, 248, 254}
\definecolor{darkblue}{RGB}{134, 134, 215}

\definecolor{ACMPurple}{cmyk}{0.55,1,0,0.15}
\definecolor{ACMBlue}{cmyk}{1,0.1,0,0.1}
\definecolor{greenfont}{RGB}{0,128,0}

\usepackage{hyperref}

\hypersetup{
    colorlinks=true,
    linkcolor=ACMPurple,
    filecolor=ACMPurple,      
    urlcolor=ACMBlue,
    citecolor=ACMPurple,
}

\newcommand\encircle[1]{%
  \tikz[baseline=(X.base)] 
    \node (X) [draw, shape=circle, inner sep=-1, fill=black, text=white, 
    minimum size=.4cm
    ] {\strut #1};%
}

\begin{document}
\bstctlcite{IEEEexample:BSTcontrol}

\title{LIDL: LLM Integration Defect Localization via Knowledge Graph-Enhanced Multi-Agent Analysis}
\author{Gou Tan$^{\orcidlink{0009-0008-6580-1470}}$,
Zilong He$^{\orcidlink{0000-0001-7963-082X}}$,
Min Li$^{\orcidlink{0009-0006-0049-361X}}$,
Pengfei Chen$^{\orcidlink{0000-0003-0972-6900}}$,
Jieke Shi$^{\orcidlink{0000-0002-0799-5018}}$,
Zhensu Sun$^{\orcidlink{0000-0001-5393-7858}}$,
Ting Zhang$^{\orcidlink{0000-0002-6001-1372}}$,\\
Danwen Chen$^{\orcidlink{0009-0006-0448-530X}}$,
Lwin Khin Shar$^{\orcidlink{0000-0001-5130-0407}}$,
Chuanfu Zhang$^{\orcidlink{0009-0007-1525-5830}}$,
and David Lo$^{\orcidlink{0000-0002-4367-7201}}$,~\IEEEmembership{Fellow, IEEE}
\thanks{Gou Tan, Min Li, and Chuanfu Zhang are with the School of Systems Science and Engineering, Sun Yat-sen University, China. Zilong He, Pengfei Chen, and Danwen Chen are with the School of Computer Science and Engineering, Sun Yat-sen University, China. E-mail: {\{tang29, hezlong, limin258, chendw9\}@mail2.sysu.edu.cn} and \{chenpf7, zhangchf9\}@mail.sysu.edu.cn.}
\thanks{Jieke Shi, Zhensu Sun, Lwin Khin Shar and David Lo are with Singapore Management University, Singapore. E-mail: {\{jiekeshi, zssun, lkshar, davidlo\}@smu.edu.sg}.}
\thanks{Ting Zhang is with Monash University, Australia. E-mail: {ting.zhang@monash.edu}.}
\thanks{Pengfei Chen and Chuanfu Zhang are the corresponding authors.}}
\markboth{IEEE TRANSACTIONS ON SOFTWARE ENGINEERING,~Vol.~00, No.~0, 2026}%
{Tan \MakeLowercase{\textit{et al.}}: LIDL}
\maketitle
\begin{abstract}
LLM-integrated software, which embeds or interacts with large language models (LLMs) as functional components, exhibits probabilistic and context-dependent behaviors that fundamentally differ from those of traditional software. This shift introduces a new category of integration defects that arise not only from code errors but also from misaligned interactions among LLM-specific artifacts, including prompts, API calls, configurations, and model outputs. However, existing defect localization techniques are ineffective at identifying these LLM-specific integration defects because they fail to capture cross-layer dependencies across heterogeneous artifacts, cannot exploit incomplete or misleading error traces, and lack semantic reasoning capabilities for identifying root causes.

To address these challenges, we propose LIDL, a multi-agent framework for defect localization in LLM-integrated software. LIDL (1) constructs a code knowledge graph enriched with LLM-related annotations that represent interaction boundaries across source code, prompts, and configuration files, (2) fuses complementary sources of error evidence to surface candidate defect locations, and (3) applies context-aware validation that uses counterfactual reasoning to distinguish true root causes from propagated symptoms. We evaluate LIDL on 146 real-world defect instances collected from 105 GitHub repositories and 16 agent-based systems. The results show that LIDL significantly outperforms seven state-of-the-art baselines in localization accuracy, improving Top-3 accuracy by 19.15\%, 21.69\%, and 21.65\% at the file, function, and line levels. Notably, LIDL achieves these gains at a lower cost than most baselines. Our localization results also improve downstream repair success by 30.77\% over the best-performing baselines.
\end{abstract}


\section{INTRODUCTION}\label{sec:INTRODUCTION}
\IEEEPARstart{R}{ecent} years have witnessed a rapid increase in the integration of Large Language Models (LLMs) into real-world software systems, resulting in a new class of applications that incorporate LLMs or invoke them programmatically as core components. We refer to this emerging category as {\it LLM-integrated software}~\cite{25BucaioniIntegrated}. Prominent examples include conversational applications like ChatGPT~\cite{25CHATGPT} and AI-assisted development tools such as GitHub Copilot~\cite{25GithubCopilot}, which streamline workflows and improve productivity, fueling a rapidly expanding market projected to reach \$36.1\text{B} by 2030~\cite{24MarketLarge} and illustrating their growing role in modern software ecosystems.





\begin{figure}[t]
    \centering
    \includegraphics[width=0.489\textwidth]{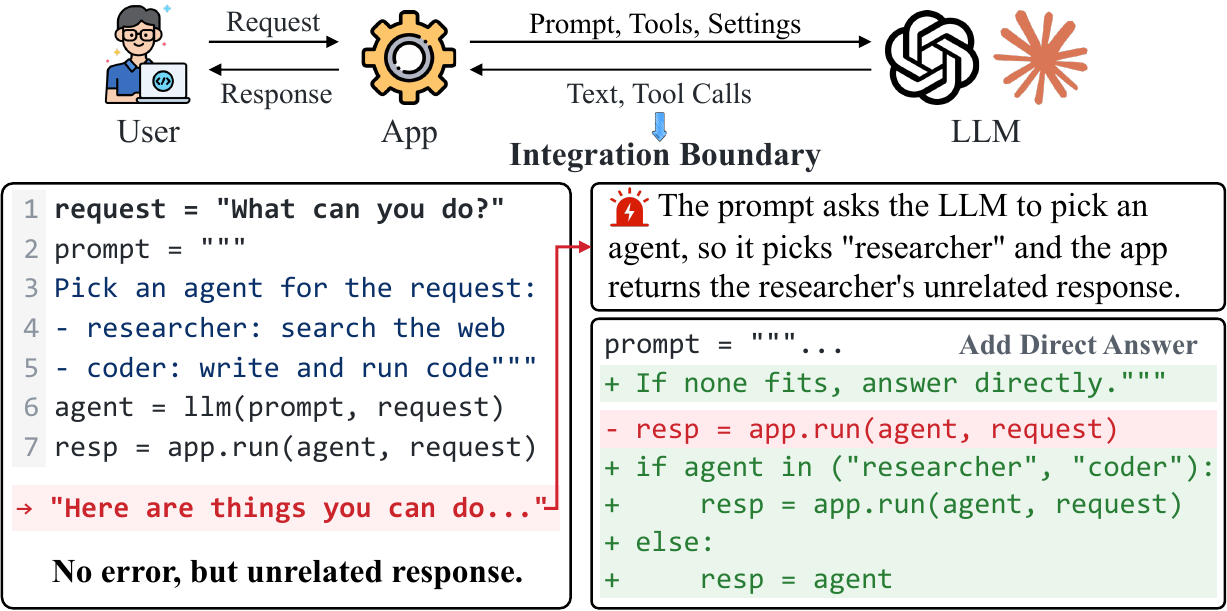}
    \caption{Overview of LLM-integrated software and an example of integration defect.}
    \label{fig:01_Integration}
    \vspace{-6mm}
\end{figure}

However, building such software reliably remains challenging. Unlike traditional software, LLM-integrated systems rely heavily on interactions with LLMs through specialized interfaces and integration modules, which introduce new complexity and, consequently, new software quality challenges~\cite{24NingDefects, 25YuSurvey}. These challenges have recently been identified as {\it LLM integration defects}~\cite{2025ShaoLLM}. Prior studies~\cite{2025ShaoLLM, Xue2025Characterization, 25CemriAgent} have shown that such defects frequently appear in modules unique to LLM-integrated software, including prompt and context management as well as LLM interface handling, such as input and output format validation. They find that 100 software projects contain integration defects that cause crashes and wrong behavior~\cite{2025ShaoLLM}, that 31.8\% of 1,026 software defects cause crashes and 30.7\% cause wrong behavior~\cite{Xue2025Characterization}, and that 44.2\% of failures in 1,642 software traces come from design defects~\cite{25CemriAgent}. Figure~\ref{fig:01_Integration} illustrates a representative case: an application invokes a researcher agent to search the web instead of describing its capabilities when asked ``What can you do?''~\cite{25ninjaIssue515}. The defect results from a prompt that requires agent selection without allowing direct answers. These defects typically arise at boundaries where software components interact with LLMs and involve indeterminate model responses and cross-component dependencies, which makes them particularly difficult for existing defect localization techniques to detect.

Concretely, existing methods face three key challenges in localizing LLM integration defects: \encircle{C1} Such defects span heterogeneous components beyond source code, including configuration files and prompts written in formats such as YAML or plain text~\cite{2025ShaoLLM}. Existing tools predominantly rely on syntactic or control-flow analysis~\cite{2013Sahalocalization, 2009Abreuspectrum} and therefore often cannot inspect these non-code artifacts, and even when they can, they struggle to construct meaningful cross-file relationships between LLM-specific artifacts and conventional code, which hinders accurate defect reasoning. \encircle{C2} This heterogeneity leads to unreliable runtime signals because error traces, which are essential to defect localization, often surface at the invocation layers of LLMs rather than at their true origins. For example, an error trace may point to a wrapper call, although the underlying issue is a misformatted prompt that should have been validated elsewhere, which renders trace-based approaches such as spectrum-based fault localization~\cite{2009Abreuspectrum} ineffective. \encircle{C3} Even when the relevant artifacts are identified, determining whether they are defective requires contextual semantic reasoning, since defects may stem from ambiguous prompt wording that triggers unintended model behavior despite a syntactically correct implementation, which existing techniques~\cite{2024XiaAgentless, 2025YangSWE, 2024ZhangAutoCodeRover} are not designed to detect.

To fill this gap, we present LIDL, a multi-agent framework for localizing LLM integration defects. The core idea of LIDL is to construct a unified knowledge representation, which is a graph of heterogeneous artifacts, and use it as the basis for LLM-driven semantic reasoning to identify defective components using evidence beyond runtime traces. LIDL operates through three coordinated agents. First, a repository graph constructor builds a knowledge graph that captures both conventional program structure and interaction points with LLMs. This graph records relationships among source code, prompts, configuration files, and other artifacts, which enables the framework to model cross-file dependencies that existing approaches cannot leverage. Second, a defect analysis agent integrates complementary evidence from direct extraction, text-based retrieval, symptom-based retrieval, annotation-based retrieval, and graph-based retrieval. This evidence fusion allows LIDL to surface plausible defect candidates even when runtime traces are incomplete or misleading. Finally, a context-aware validator applies counterfactual reasoning to assess whether correcting a suspected defective component would resolve the defect, allowing LIDL to distinguish true root causes from secondary effects and rank candidates using contextual semantics.

To evaluate LIDL, we constructed a benchmark covering four categories of LLM integration defects, consisting of 146 real-world defects collected from 105 GitHub repositories and 16 agent-based systems. Since no existing techniques specifically target LLM integration defects, we compared LIDL with seven state-of-the-art repository-level defect localization methods: SWE-agent~\cite{2025YangSWE}, Agentless~\cite{2024XiaAgentless}, AutoCodeRover~\cite{2024ZhangAutoCodeRover}, LocAgent~\cite{Chen2025LocAgent}, Codex~\cite{2026OpenAICodex}, and RepoGraph-enhanced approaches (SWE-agent$^*$ and Agentless$^*$)~\cite{2024OuyangRepoGraph}, which augment the original methods with repository-level code structure graphs. This comparison quantifies the performance gap when applying traditional defect localization to LLM-integrated software. Experimental results show that LIDL significantly outperforms all baselines at the file, function, and line levels across the evaluated backbone LLMs, improving Top-3 accuracy by 19.15\%, 21.69\%, and 21.65\%. In addition to accuracy gains, LIDL achieves lower costs than most baselines, averaging \$0.107 per localization task. Our localization results also improve downstream repair success by 30.77\% over the best-performing baselines. Finally, ablation studies show how the core components affect localization performance.

The main contributions of this work are as follows:

\begin{itemize}[leftmargin=*]
\item We are the first to propose integrating a code knowledge graph with LLM-based semantic reasoning to address the unique challenges of localizing LLM integration defects.
\item We implement LIDL, a multi-agent framework for defect localization of LLM-integrated software, combining knowledge graph, multi-source evidence fusion, and counterfactual validation in a unified workflow.
\item We evaluate LIDL on 146 real-world defect instances and show that it significantly outperforms seven state-of-the-art baselines in localization accuracy and downstream repair success with low cost. All benchmark data and our implementation are publicly available at: \url{https://github.com/IntelligentDDS/LIDL}.
\end{itemize}

The rest of the paper is organized as follows. Section~\ref{sec:PRELIMINARIES} introduces the background and presents our analysis of LLM integration defects. Section~\ref{sec:METHODOLOGY} introduces the LIDL framework. Section~\ref{sec:EVALUATION} presents the experimental results and evaluation. Section~\ref{sec:DISCUSSION} discusses implications, limitations, and future work, followed by threats to validity in Section~\ref{sec:THREATS} and conclusions in Section~\ref{sec:CONCLUSION}.

\section{PRELIMINARIES}\label{sec:PRELIMINARIES}
\subsection{LLM-integrated Software} 
LLM-integrated software refers to applications that embed or invoke LLMs as functional components, enabling capabilities such as natural language understanding~\cite{25CHATGPT} and code generation~\cite{25GithubCopilot}. In this type of software, LLMs participate in content generation or decision making, meaning that software behavior depends not only on code but also on model outputs, prompts, and runtime context. As a result, LLM-integrated software exhibits probabilistic and context-sensitive behavior, which distinguishes it from traditional software.

In practice, LLM-integrated software typically follows one of three architectural patterns: (1) {\bf Direct LLM Invocation}, where applications call LLMs through APIs or local models, for example ChatGPT clients~\cite{25CHATGPT} and code completion tools~\cite{25GithubCopilot}; (2) {\bf Retrieval-Augmented Generation (RAG)}, which improves response quality by retrieving external knowledge bases at runtime, as seen in systems such as Dify~\cite{2025DifyLLM} and PrivateGPT~\cite{2025PrivateGPTLLM}; (3) {\bf Agent-based Architectures}, which coordinate multi-step reasoning and tool execution by combining LLMs with memory modules and planning mechanisms, such as AutoGen~\cite{2023WuAutoGen} and MetaGPT~\cite{2024HongMetaGPT}.

These architectural patterns are commonly implemented using frameworks such as LangChain~\cite{2022LangChainBuild} and LlamaIndex~\cite{2022LiuLlamaIndex}. These frameworks provide standardized abstractions for prompt management, context control, and tool invocation, but also introduce additional integration complexity that leads to new failure modes.

\subsection{LLM Integration Defects}\label{subsec:IntegrationDefect}
Traditional software defects originate from issues in the code itself, such as incorrect implementations, API misuse, or faulty dependency handling~\cite{2025ChangBridging, 25NiuWhen}. In contrast, LLM integration defects often arise from interactions between code and LLMs rather than code errors alone. Such defects may stem from prompt phrasing, context management, model responses, configuration settings, or the dynamic behavior of tool–LLM orchestration rather than deterministic computation. Following prior studies~\cite{2025ShaoLLM,25RahardjaCan,24NingDefects}, we categorize these LLM integration defects into four primary groups:
\begin{itemize}[leftmargin=*]
\item \textbf{Prompt and Context Management}: Defects caused by unclear, incomplete, or improperly maintained prompt or context information, leading to undesired or inconsistent model responses.
\item \textbf{LLM Interface Management}: Defects stemming from violations of LLM input/output requirements, such as unvalidated prompt format, mismatched output schema, or exceeding token and context limits.
\item \textbf{Tool Integration Management}: Defects occurring when LLM-driven components interact with external tools, including incorrect invocation parameters, misconfigured dependencies, or tool execution failures.
\item \textbf{LLM System Management}: System-level defects involving configuration, resource management, deployment, or security, such as misconfigured API keys, throttling, access control issues, or runtime resource constraints.
\end{itemize}

\begin{figure*}[t]
    \centering
    \includegraphics[width=\textwidth]{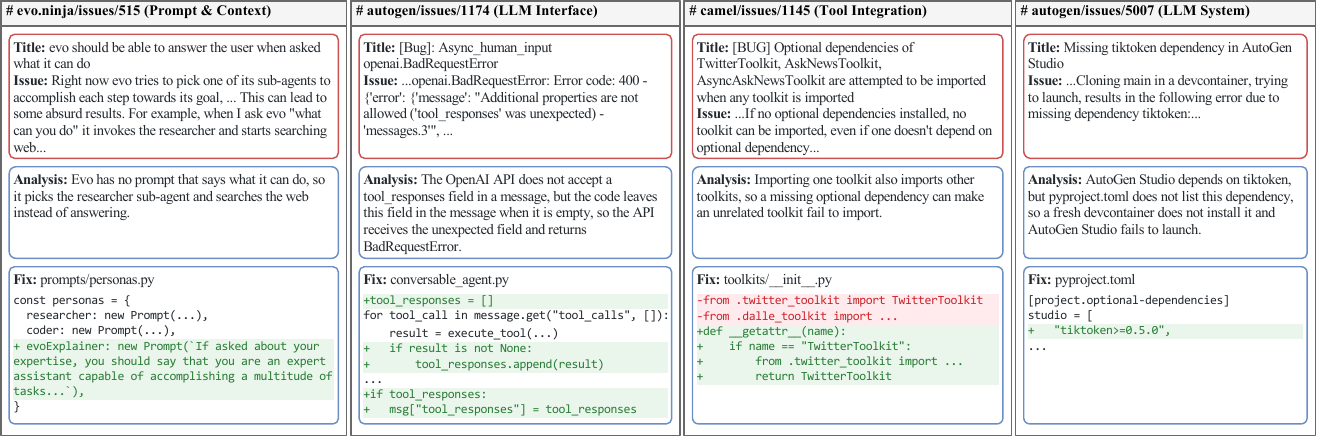}
    \caption{Representative defect cases across four LLM integration defect categories~\cite{25ninjaIssue515, 25autogenIssue1174, 25camelIssue1145, 25autogenIssue5007}.}
    \label{fig:4defect_cases}
    \vspace{-2mm}
\end{figure*}

To better illustrate these defect characteristics, we analyze the 146 defects in our evaluation dataset (\S\ref{subsec:Experiment Setup}), and present four representative cases (Fig.~\ref{fig:4defect_cases}), one for each category. These cases are selected from GitHub issues in widely used LLM-integrated software and illustrate how LLM integration defects differ from traditional software defects.

\textbf{Case 1: evo.ninja issue \#515~\cite{25ninjaIssue515} (Prompt and Context).}
When users asked ``what can you do?'', the system triggered a researcher sub-agent to perform a web search rather than providing a self-description. The observed behavior pointed to the agent selection logic, but the root cause was a missing introspection prompt template. The fix added a prompt for describing the system's capabilities. Identifying the root cause required reasoning about query intent rather than following the trace.

\textbf{Case 2: autogen issue \#1174~\cite{25autogenIssue1174} (LLM Interface).}
The system failed with \texttt{openai.BadRequestError}: ``tool\_responses was unexpected''. The trace indicated the API invocation, but the defect was improper handling of empty tool response lists, which caused an unsupported field to be sent. The fix ensured ``tool\_responses'' is only included when tool output exists.

\textbf{Case 3: camel issue \#1145~\cite{25camelIssue1145} (Tool Integration).}
Importing any toolkit could cause \texttt{ModuleNotFoundError} because the initializer eagerly imported other toolkits that required optional dependencies. The fix restructured initialization to delay dependency resolution. This defect spans registry files, dependency declarations, and loading code.

\textbf{Case 4: autogen issue \#5007~\cite{25autogenIssue5007} (LLM System).}
AutoGen Studio failed to launch due to missing \texttt{tiktoken}. The trace identified the import chain, but the root cause was the absence of this dependency in \texttt{pyproject.toml}. The defect originated in the configuration file rather than executable code.


\subsection{Challenges for LLM Integration Defect Localization}\label{subsec:Challengesfor}
Traditional defect localization aims to identify suspicious code regions responsible for software failures~\cite{2016WongLocalization}. Existing approaches include four types. Spectrum-based fault localization (SBFL)~\cite{2009Abreuspectrum} ranks code elements by their correlation with test failures. Mutation-based fault localization (MBFL)~\cite{2015Papadakismutation} injects faults to observe behavioral changes. Information retrieval-based approaches~\cite{12ZhouBugLocator, 2013Sahalocalization,10.1109/ASE51524.2021.9678546} match defect reports to source code using textual similarity. Learning-based approaches~\cite{2019LiDeepFL, 2021LouBoosting, 2022QiuDeep, 2022MengImproving, 2023ZhangContext} apply machine learning to learn defect patterns from code and execution data.

However, these methods face limitations when applied to LLM-integrated software, where failures may involve model interactions and non-code artifacts. Different methods face different challenges in this setting. Information retrieval-based methods often fail because LLM defects involve inconsistent terminology and occur in non-code artifacts. SBFL and MBFL rely on deterministic test reproduction, but LLM failures may occur without crashes, require conversational context, or vary across executions. Learning-based models require training on large and representative datasets, but we currently lack a large dataset of prompts, LLM interfaces, and tool orchestration patterns that appear in LLM integration defects, which limits their applicability.

Recent work has explored LLM-based defect localization~\cite{2024WangOpenHands,2025YangSWE,2024MaLingmaAgent,2024AntoniadesSearch,Reddy2026SWERank,Sepidband2026RGFL,Seddik2026ARISE}. Existing techniques fall into three categories.
Agent-based approaches, for example, OpenHands~\cite{2024WangOpenHands}, SWE-agent~\cite{2025YangSWE}, and Codex~\cite{2026OpenAICodex}, perform repository-level reasoning by iteratively exploring, executing, and editing files.
Hierarchical approaches, for example Agentless~\cite{2024XiaAgentless}, BugCerberus~\cite{2025ChangBridging}, and FlexFL~\cite{2025XuFlexFL}, progressively narrow search scopes.
Repository-structured approaches extend LLM reasoning with code skeletons or repository graphs, for example, AutoCodeRover~\cite{2024ZhangAutoCodeRover}, RepoGraph~\cite{2024OuyangRepoGraph}, LocAgent~\cite{Chen2025LocAgent}, and CodexGraph~\cite{2024LiuCodexGraph}.
However, these approaches may miss defects in configuration or prompt layers, often traversing files that are irrelevant to the defect. Repository graph approaches, although they have better capability in inspecting files, do not distinguish LLM-related artifacts from code and often omit non-code files, such as prompt templates or YAML/TOML configurations.

\begin{figure*}[t]
    \centering    
    \includegraphics[width=\textwidth]{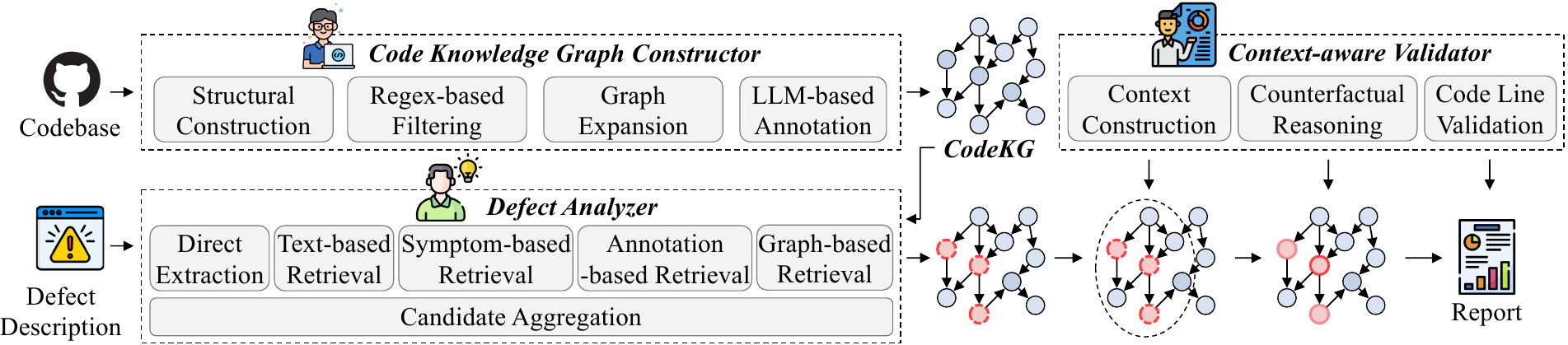}
    \caption{The architecture of LIDL.}
    \label{fig:1framework}
\end{figure*}

\noindent\textbf{Summary of Challenges.}
Based on the analysis above, as well as the representative cases in Section~\ref{subsec:IntegrationDefect}, we elaborate three challenges that must be addressed for localizing LLM integration defects:
\begin{itemize}[leftmargin=*]
\item \textbf{C1: Heterogeneous Components.} Cases~3 and~4 show that defects can involve toolkit initialization code and configuration files such as \texttt{pyproject.toml}. Existing tools may overlook non-code artifacts and their relationships with source code.

\item \textbf{C2: Unreliable Runtime Signals.} Cases~1 and~2 show that observed behavior and error traces reveal agent selection and API invocation, while the actual defects reside in prompt templates and response handling. Trace-based methods fail when runtime signals are misleading.

\item \textbf{C3: Contextual Semantic Reasoning.} Case~1 shows that the query ``what can you do?'' triggers web search instead of self-description due to missing prompt wording. Identifying such defects requires reasoning about prompt semantics rather than code structure.
\end{itemize}

Neither traditional nor existing LLM-based methods are sufficient to address these challenges, which motivates our approach.





\section{METHODOLOGY}\label{sec:METHODOLOGY}
In this paper, we propose LIDL, a framework for localizing defects in LLM-integrated software by combining structural repository knowledge with LLM-based reasoning. As shown in Fig.~\ref{fig:1framework}, LIDL adopts a multi-agent architecture to address the challenges summarized in Section~\ref{sec:PRELIMINARIES}. The framework consists of three agents: a \textit{Code Knowledge Graph Constructor}, a \textit{Defect Analyzer}, and a \textit{Context-aware Validator}. Given a codebase, the \textit{Code Knowledge Graph Constructor} builds a knowledge graph that captures both conventional program structure and interaction points with LLMs. The graph records relationships among source code, prompts, configuration files, and other artifacts, which enables modeling of cross-file dependencies. It also annotates files with their functional roles in the LLM workflow, such as prompt template construction. During this process, LIDL uses a pattern library of LLM-specific keywords collected from popular frameworks to select files for annotation.

Using the constructed graph, the \textit{Defect Analyzer} retrieves and prioritizes suspicious files based on defect descriptions, runtime signals, and pattern-based semantic reasoning, and it gradually narrows the search space to candidate code lines by fusing heterogeneous evidence rather than relying on a single signal source. Finally, the \textit{Context-aware Validator} applies counterfactual reasoning by asking an LLM to use code context to assess whether candidate files can explain the defect symptoms and whether correcting faulty logic in them could eliminate the symptoms. This assessment guides the reranking of candidate code lines. The validator then checks the complete source of the selected files to identify the code lines most likely to contain the defect.

\begin{figure*}[t]
    \centering
    \includegraphics[width=\textwidth]{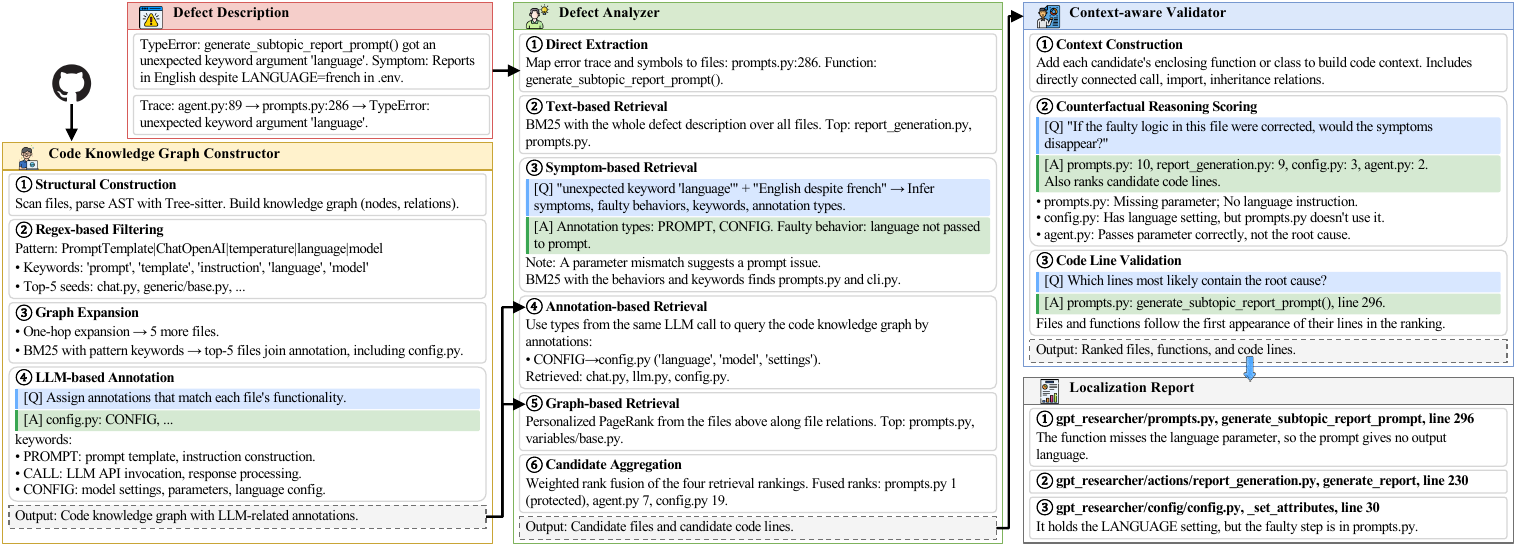}
    \caption{End-to-end running example of LIDL on a real defect from gpt-researcher~\cite{25researcherIssue1027}.}
    \label{fig:3LocalizationReport}
    \vspace{-3mm}
\end{figure*}

\vspace{0.05em}
\noindent
\textbf{Running Example.} We use a real defect from gpt-researcher~\cite{25researcherIssue1027} to illustrate LIDL's workflow (Fig.~\ref{fig:3LocalizationReport}). The defect raises a \texttt{TypeError}: the function \texttt{generate\_subtopic\_report\_prompt()} receives an unexpected \texttt{language} parameter. The system outputs English even when \texttt{LANGUAGE="french"} is specified in the configuration file. The reported symptoms suggest checking how the language setting is passed to prompt construction. We reference this example in subsequent sections to show how each component processes this defect.

\vspace{0.05em}
\noindent
\textbf{Notation.} We represent a repository as a code knowledge graph $G = (V, E)$, where nodes $V$ include files, classes, functions, and other code entities, and edges $E$ capture relationships such as \textit{contain}, \textit{call}, \textit{import}, and \textit{extend} (Table~\ref{tab:EdgeTypes}). We use $V_f \subseteq V$ to denote file nodes (FILE and TEXTFILE in Table~\ref{tab:NodeTypes}). We use $D$ to denote the defect description provided as input. The pattern library $P$ stores regular expression patterns associated with the LLM-related annotation types (Table~\ref{tab:LLMannotations}).

\begin{table}[t]
\belowrulesep=0pt
\aboverulesep=0pt
\renewcommand{\arraystretch}{1.5}
\centering
\caption{Node types in the code knowledge graph.}
\label{tab:NodeTypes}
\setlength{\tabcolsep}{0.005\textwidth}
\begin{tabular}{>{\centering\arraybackslash}m{0.1\textwidth}|>{\raggedright\arraybackslash}m{0.368\textwidth}}
\bottomrule
\textbf{Type} & \multicolumn{1}{c}{\textbf{Description}} \\ \midrule
REPO & Virtual root node representing the entire repository. \\ \hline
PACKAGE & Virtual node representing a directory in the file system. \\ \hline
FILE & Source code files (e.g., .py, .java). \\ \hline
TEXTFILE & Configuration and template files (e.g., .yaml, .jinja2). \\ \hline
CLASS & Class definitions in object-oriented programming. \\ \hline
FUNCTION & Function and method definitions. \\ \hline
ATTRIBUTE & Global variables and class attributes. \\
\toprule
\end{tabular}
\end{table}
\begin{table}[t]
\belowrulesep=0pt
\aboverulesep=0pt
\renewcommand{\arraystretch}{1.5}
\centering
\caption{Edge types in the code knowledge graph.}
\label{tab:EdgeTypes}
\setlength{\tabcolsep}{0.005\textwidth}
\begin{tabular}{>{\centering\arraybackslash}m{0.1\textwidth}|>{\raggedright\arraybackslash}m{0.368\textwidth}}
\bottomrule
\textbf{Type} & \multicolumn{1}{c}{\textbf{Description}} \\ \midrule
CONTAIN & Hierarchical containment: a repository contains packages, a package contains files, a file contains classes or functions, and a class contains methods or attributes. \\ \hline
CALL & Function invocation: one function calls another. \\ \hline
IMPORT & Dependency: one file imports another file or configuration. \\ \hline
EXTEND & Class inheritance: one class extends another. \\
\toprule
\end{tabular}
\end{table}
\begin{table}[t]
\belowrulesep=0pt
\aboverulesep=0pt
\renewcommand{\arraystretch}{1.5}
\centering
\caption{LLM-related annotations for files.}
\label{tab:LLMannotations}
\setlength{\tabcolsep}{0.005\textwidth}
\begin{tabular}{>{\centering\arraybackslash}m{0.12\textwidth}|>{\raggedright\arraybackslash}m{0.358\textwidth}}
\bottomrule
\textbf{Type} & \multicolumn{1}{c}{\textbf{Description}} \\
\midrule
LLM\_PROMPT & Prompt template construction and formatting (e.g., system, user, prompt, instruction). \\ \hline
LLM\_CALL & LLM API invocations and method calls (e.g., ChatOpenAI.agenerate(), model.invoke()). \\ \hline
LLM\_CONFIG & LLM configuration and parameter settings (e.g., model\_name, temperature, api\_key). \\ \hline
LLM\_TOOL & Tool registration and function definitions (e.g., @tool and register\_tool calls). \\ \hline
LLM\_MEMORY & Conversation history and vector storage management (e.g., ConversationBufferMemory, VectorStore). \\
\toprule
\end{tabular}
\end{table}

\subsection{Code Knowledge Graph Constructor}
\label{sec:graph}
As discussed earlier, the first challenge lies in representing heterogeneous components and their interactions in LLM-integrated systems. To address this, the \textit{Code Knowledge Graph Constructor} builds a structural–semantic hybrid repository representation that captures not only conventional program structure but also the operational roles of artifacts involved in LLM workflows. Existing repository graphs, for example, RepoGraph\cite{2024OuyangRepoGraph} and CodexGraph~\cite{2024LiuCodexGraph}, primarily model syntax-level entities such as functions, classes, and imports, which makes them insufficient for cases in which defects originate from prompts, configuration files, or model invocation logic.

To overcome these limitations, our approach extends repository modeling along two dimensions. First, it expands node types beyond source code to include non-code artifacts, such as configuration files, prompt templates, and tool-binding specifications, which influence LLM execution and behavior. Second, it assigns semantic role annotations, for example \texttt{LLM\_CALL} and \texttt{LLM\_CONFIG}, based on the functional purpose of each artifact within the execution pipeline. These annotations allow us to distinguish LLM-specific components from conventional logic.

The resulting code knowledge graph $G$ includes both structural relations, such as call chains, imports, and file inclusion paths, and LLM-related annotations stored as file node attributes. The relations connect files across components, and the annotations show what each file does, so the Defect Analyzer can use both. To support efficient downstream processing, $G$ is indexed using global lookup tables, which enables $O(1)$ artifact retrieval and scalable traversal during localization.

\subsubsection{Structural Construction}
Building on our definition of the code knowledge graph $G$, we construct nodes representing code entities and edges representing their relationships.
Each node $v \in V$ stores its type, name, file path, and source text.
Each edge $e = (v_i, v_j) \in E$ connects source $v_i$ to target $v_j$ with a relationship type.

The constructor scans the repository and parses code files to build the graph. Before parsing, it applies several filtering rules: skipping common directories (e.g., \texttt{\_\_pycache\_\_}, \texttt{.git}), including files with extensions relevant to LLM-integrated software (e.g., \texttt{.py}, \texttt{.yaml}, \texttt{.json}, \texttt{.jinja2}, \texttt{.txt}), excluding auto-generated files and \texttt{\_\_init\_\_.py} files without substantive content, and skipping hidden files. 

 Source code files supported by Tree-sitter~\cite{2025Treesitter} are then parsed to extract the information required to build the graph.
Specifically, a graph $G$ consists of 7 node types and 4 edge types, respectively shown in Table~\ref{tab:NodeTypes} and Table~\ref{tab:EdgeTypes}.
The nodes include not only syntactic units, such as classes and functions, but also configuration files and prompt templates that influence LLM behavior.
By mapping these node and edge types to the extracted repository artifacts, we build a graph that records structural relations among code entities and files. We resolve call targets using symbols within the file, receiver constraints, and import information, and assign each candidate relation a confidence equal to the reciprocal of the number of candidate targets.


\begin{figure}[t]
    \centering
    \includegraphics[width=0.489\textwidth]{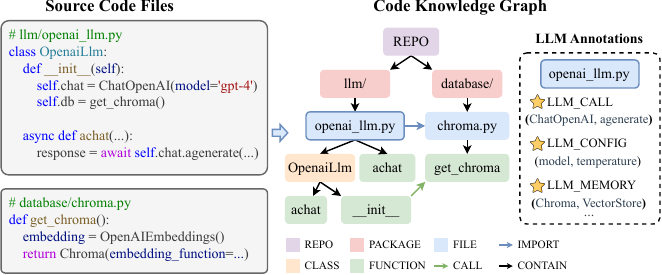}
    \caption{Example of code knowledge graph with LLM annotations. The code is from an open-source LLM application~\cite{25RealCharCreate}.}
    \label{fig:2codeKG}
    \vspace{-4mm}
\end{figure}




\subsubsection{LLM-related Annotation}\label{sec:semantic-annotation}
While the structural graph provides dependency and relationship information, it does not reveal how each artifact participates in the LLM workflow. To bridge this gap, LIDL applies LLM-related annotation to label candidate files with operational roles (e.g., \texttt{LLM\_CALL}, \texttt{LLM\_CONFIG}, \texttt{LLM\_MEMORY}), which enables downstream reasoning over LLM-specific behaviors. The annotation process starts with regex-based filtering to select initial files. Graph expansion finds directly connected files, which are then filtered using BM25 based on text relevance. An LLM assigns annotations to the selected files.

\vspace{0.1cm}
\noindent
\textbf{Regex-based Filtering.} Since a subset of repository files relates to LLM behavior, this step filters irrelevant files before invoking expensive LLM reasoning. The constructor uses a pattern library $P$, where a pattern refers to a keyword or phrase commonly associated with LLM-related functionality (e.g., \texttt{ChatOpenAI}, \texttt{system\_prompt}, \texttt{@tool}) and is stored in regex form for matching. Each annotation type in $P$ contains patterns manually collected from widely used LLM frameworks.

For each file $v_f \in V_f$, the constructor counts the distinct patterns matched and the total pattern matches. Dividing the distinct pattern count by the total number of patterns in $P$ gives coverage. Density is the number of pattern matches per thousand characters. We rank files using the fixed heuristic score $0.7\times\mathrm{coverage}+\min(\mathrm{density}/10,0.3)$. The top-$k_s$ ranked files serve as analysis seeds for subsequent graph expansion.

\vspace{0.1cm}
\noindent
\textbf{Graph Expansion.}
Since LLM-related logic may be distributed across multiple interacting files,
we expand the initial seed set 
by one hop through the code knowledge graph.
Specifically, starting from each seed node, we retrieve all nodes
that are reachable within one edge hop in the graph, where an edge represents
a structural or dependency relation defined in Section~\ref{sec:graph}. The expansion follows both incoming and outgoing relations. To focus on concrete artifacts, we retain file nodes (i.e., \textsc{FILE} and \textsc{TEXTFILE}) and map retrieved functions or classes to their containing files.

 We then use the LLM-related pattern keywords from the regex-based filtering stage as the query. BM25~\cite{robertson2009probabilistic} measures the lexical relevance between this query and the text of each retrieved file, and we rank the files accordingly. Finally, the top-$k_e$ ranked files are merged with the original seeds, forming the final set of files for annotation.


\vspace{0.1cm}
\noindent
\textbf{LLM-based Annotation.}\label{sec:LLM-based Annotation}
The constructor invokes an LLM to annotate the filtered files, assigning each file one or more labels from the five annotation types in Table~\ref{tab:LLMannotations}. Files are processed in batches within the model's context limit, with oversized files split as needed. The LLM also returns a short phrase explaining why each annotation applies. 

The LLM also extracts keywords from the source code to capture project-specific terms that the predefined patterns may miss. To reduce hallucination, extracted keywords are validated against the source code, and any keyword not literally present in the file is discarded. Within each annotation record, duplicate keywords are removed, and when one keyword is a prefix of another, the shorter keyword is retained. Validated keywords are converted to regex patterns by escaping special characters to match each retained keyword as written. After annotation, we use these patterns to match additional files in the repository for the current defect instance. These files receive the annotation types associated with the matched keywords, and we call the matched keywords supplementary keywords.

The graph is enhanced by attaching an LLM-related annotation attribute to each annotated file node. This attribute stores the annotation type, descriptive phrase, and validated keywords, enabling the Defect Analyzer to retrieve files by querying annotation labels directly. Fig.~\ref{fig:2codeKG} illustrates an annotated graph in which \texttt{openai\_llm.py} is labeled with \texttt{LLM\_CALL} (for ChatOpenAI and agenerate), \texttt{LLM\_CONFIG} (for model and temperature settings), and \texttt{LLM\_MEMORY} (for Chroma and VectorStore). In our running example, the constructor annotates \texttt{config.py} with \texttt{LLM\_CONFIG}. \texttt{prompts.py} has no annotation, and the analyzer finds it through other retrieval sources.

    
    

\subsection{Defect Analyzer}
 To address missing or misleading error traces, this agent identifies candidate files through complementary retrieval methods. It takes the defect description $D$ and the graph $G$ constructed by the \textit{Code Knowledge Graph Constructor} as input and outputs candidate files and candidate code lines. The defect description contains the text of a bug report, such as a GitHub issue, with error messages, error traces, and observed symptoms (e.g., \texttt{BadRequestError}: ``tool\_responses was unexpected''). The analyzer uses direct extraction to identify files and code symbols explicitly mentioned in the defect description. It then uses retrieval methods that match text and extend the search through the code knowledge graph. Text-based retrieval matches the whole defect description with file text, and symptom-based retrieval matches the symptoms and keywords that an LLM infers from the description with file text. Annotation-based retrieval extends the search through the LLM-related annotations of files, and graph-based retrieval extends it along the relations between files. The analyzer uses a single structured LLM call to extract the information needed for symptom-based retrieval and annotation-based retrieval. Candidate aggregation then fuses the results and narrows the candidates to code lines.

\subsubsection{Direct Extraction}
This component adopts the most straightforward strategy, i.e., locating candidate files based on explicit signals in the defect description, including error traces and file references. Specifically, we map file paths in error traces, explicit file mentions, and code symbols such as function and class names to repository files. We keep a mapping only when the file name or code symbol identifies one file. We keep all mapped files without further filtering, because they come directly from the defect description.

\subsubsection{Text-based Retrieval}
This component matches the complete defect description with the text of every file in the repository. It uses BM25 to produce a text-relevance ranking, and it can include files that the other methods miss.

\subsubsection{Symptom-based Retrieval}
This component infers candidate files by reasoning about defect symptoms, even when no file paths appear in the description. The analyzer prompts an LLM to extract failure stages, symptoms, possible faulty behaviors, related code symbols, retrieval keywords, predicted annotation types, and fault hypotheses from the defect description. Each fault hypothesis describes a possibly faulty operation, an involved symbol, a wrong value, and an expected constraint.

 The analyzer then matches the related code symbols with repository symbols and uses the faulty behaviors and retrieval keywords to search file paths, symbol names, and source code with BM25. Behavior matching evidence refers to word matches for an inferred behavior, checked separately in the defect description and repository source code. These matches do not establish that a file contains a defect. The analyzer ranks the retrieved files first by exact symbol matches. It then considers behavior matching evidence in the defect description, behavior matching evidence in the source code, the number of matched words, and BM25 scores, in that order.
It retains files with exact symbol matches, behavior matching evidence, or multiple matched words. To reduce missed entry files, configuration files, and top-level modules, the analyzer also retains a small number of files near the repository root with positive BM25 scores.

\subsubsection{Annotation-based Retrieval}
This component retrieves files whose LLM-related annotations match the predicted annotation types. The LLM call used for symptom-based retrieval also predicts which annotation types from Table~\ref{tab:LLMannotations} are likely involved. For example, ``vague prompt'' or ``unexpected output'' suggests \texttt{LLM\_PROMPT}. ``API error'' or ``timeout'' suggests \texttt{LLM\_CALL} or \texttt{LLM\_CONFIG}, and ``missing context'' suggests \texttt{LLM\_MEMORY}. The analyzer then selects from the graph all files whose annotations match the predicted types. For files found by the constructor through supplementary keywords, it also matches the annotation types associated with those keywords.

Because this may return many files, the analyzer compares words in the defect description with each file's annotation descriptions, validated keywords, and supplementary keywords for the current defect instance. It ranks files first by the number of matched words, then by the number of supplementary keywords associated with each file, and finally by file path.
The analyzer retains files with matched words or, if none match, the highest-ranked file with a matching annotation type.

\subsubsection{Graph-based Retrieval}
This component uses the relations in the code knowledge graph to reach related files in other components. Personalized PageRank~\cite{Zhao2025Completion} starts from the files found by direct extraction, text-based retrieval, symptom-based retrieval, and annotation-based retrieval, and produces a structural-relevance ranking. Relevance propagates in both directions with weights based on relation type, direction, and confidence. The graph used for file expansion excludes relations whose call targets are ambiguous, while the full graph used for retrieving code entities retains them with reduced weights.

\subsubsection{Candidate Aggregation}

\textbf{File-level fusion.} The analyzer combines file rankings from different sources into a unified candidate file list. These retrieval methods score files differently, so their scores cannot be compared directly. The analyzer therefore uses weighted reciprocal rank fusion~\cite{Liu25Exp4Fuse} to combine file rankings from symptom-based retrieval, annotation-based retrieval, text-based retrieval, and graph-based retrieval. The file score is $S_F(f)=\sum_r w_r/(k+\operatorname{rank}_r(f))$, where $r$ indexes the rankings containing file $f$, $w_r$ is the source weight, and $k$ is a fixed smoothing constant. Files identified by direct extraction receive an additional score. The analyzer keeps the highest-scored files as candidate files and always keeps the files from direct extraction. The analyzer merges files with identical contents while retaining their path aliases.

\textbf{Code-level fusion.} The analyzer uses Tree-sitter to extract complete assignment, condition, call, return, exception, import, and declaration statements from the candidate files. We represent candidate code lines as statements with line ranges and link them to their files and enclosing functions, classes, or modules. Long statements are split into consecutive parts. The retrieval score of a code entity, such as a function or class, combines text relevance, graph relevance, direct-extraction evidence, and annotation-based retrieval evidence. The analyzer initially ranks these candidate code lines using the retrieval scores of their files and enclosing code entities. This ranking also considers code matches to related symbols, fault hypothesis keywords, possible faulty operations, and words in the defect description.

To control the input size for validation, the analyzer first retains candidate code lines with strong fault hypothesis evidence while covering candidate files and highly ranked retrieval results. Strong evidence requires a match to the hypothesis symbol and either an operation word match or a minimum proportion of matched hypothesis words. It then uses maximal marginal relevance~\cite{Chen2025SaraCoder} to fill the remaining candidate slots, favoring relevant code while reducing repeated selection from the same file, the same code block, or similar code. The analyzer passes the selected candidates with their line ranges and code context, together with the unified candidate file list, to the \textit{Context-aware Validator}.

\subsection{Context-aware Validator}
 A highly ranked candidate may contain the root cause or may only propagate or expose the defect. To address C3 (contextual semantic reasoning), the \textit{Context-aware Validator} checks the candidates produced by the analyzer. The validator first adds code context to each candidate, then uses an LLM to score candidate files and rerank candidate code lines, and finally reads the complete source code to identify specific code lines.

\subsubsection{Context Construction}


\begin{figure}[t]
\begin{tcolorbox}[title={Validator Prompt (Selected Parts)},boxsep=1pt,left=5pt,right=5pt,top=1pt,bottom=1pt,toptitle=0pt,bottomtitle=0pt,width=\linewidth,colback=white,colframe=black!25,boxrule=0.4pt,colbacktitle=black!15,coltitle=black,fonttitle=\small\bfseries]
\fontsize{8}{8.5}\selectfont\raggedright
\setlength{\parindent}{0pt}
\setlength{\parskip}{2pt}
Your task is to assign every supplied candidate file a counterfactual score and rank every supplied statement anchor by root-cause likelihood.
\par
\textbf{File scoring}
\par
Do ask for each file: if its faulty operation were correctly repaired while all other code remained unchanged, would the reported defect disappear?
\par
Do assign every supplied file an integer score from 1 to 10, where 8-10 means a likely root cause, 6-7 means a contributing or propagating file, and 1-5 means a symptom or weakly related file.
\par
Do infer the smallest plausible correction only to judge whether the file contains the root cause. Do not generate, apply, or execute a patch.
\par
\textbf{Line ranking}
\par
Do rank every supplied Statement anchor ID exactly once from most to least suspicious.
\par
Do rank root-cause decisions and transformations above propagation calls, exception sites, definitions, documentation, tests, and nearby context.
\par
Do keep file and Statement decisions hierarchical. Rank a Statement in a lower-scored file above Statements in a higher-scored file only when its supplied evidence gives stronger root-cause support.
\par
\textbf{Input}
\par
Do use only the issue, fault hypotheses, retrieval sources, semantic annotations, confidence, source context, and CodeKG relations supplied in the input.
\par
Do treat traces and graph relations as evidence rather than proof.
\par
Do not invent files, Statements, evidence, or IDs.
\par
\textbf{Output}
\par
Do return exactly one valid JSON object shaped as
\par
\texttt{\{\allowbreak"scored\_files":\{"F1":10,"F2":1\},}
\par
\texttt{\phantom{\{}"ranked\_locations":["L2","L1"]\}},
\par
containing every supplied file ID and Statement ID exactly once, with no reasons, Markdown, or additional fields.
\end{tcolorbox}
\caption{Example of the file scoring and line ranking prompt.}
\label{fig:file-scoring-prompt}
\end{figure}

 Assessing whether correcting a candidate file could eliminate the defect requires examining the file with its code context, since LLM integration defects often arise from interactions across prompt files, API calls, and configuration dependencies. Therefore, the validator first identifies the function, class, or module that contains each candidate. It then extracts the call, import, and inheritance relations directly connected to the candidate file and its related entities from the graph. The resulting code context contains the candidate source, its enclosing function or class, the LLM-related annotations and retrieval source of its file, and the directly connected graph relations. This code context enables the LLM to distinguish code that contains the root cause from code that propagates or exposes the defect.

\subsubsection{Counterfactual Reasoning Scoring}
LLM integration defects may involve interactions between natural language content and program logic. Defect localization therefore requires semantic reasoning with code context, and the validator uses counterfactual reasoning to assess candidate files. In our setting, counterfactual reasoning asks a hypothetical question for each candidate: \emph{``If this file were correctly fixed, would the defect still occur?''} Files whose modification is semantically likely to eliminate the failure receive higher scores, whereas those that reflect only propagated effects or surface symptoms receive lower scores. The validator provides the LLM with the defect description, fault hypotheses, candidate files, candidate code lines, and their code contexts. In the same call, the LLM assesses candidate files and ranks candidate code lines. For each candidate file, the LLM assesses whether the file can explain the defect symptoms and whether the defect would disappear if the faulty logic in that file were corrected. The LLM assigns each candidate file a counterfactual score from 1 to 10 based on this assessment. Higher values indicate a stronger likelihood that the file contains the root cause. Lower scores indicate that the file is more likely to propagate or expose the defect or be weakly related to it. Fig.~\ref{fig:file-scoring-prompt} shows selected parts of the prompt for joint file scoring and candidate code line ranking. The LLM ranks candidate code lines using both the candidate source and the file assessment, which guides but does not directly determine the ranking. A candidate code line in a lower-scored file can rank above lines in a higher-scored file if it directly constructs a wrong parameter, performs a wrong conversion, or omits a required check. This stage is a hypothetical judgment by the LLM, not a formal causal analysis. It assesses the given candidates without expanding the candidate set, generating patches, or modifying or executing the code.

\subsubsection{Code Line Validation}

 The previous step ranks candidate code lines using code snippets, and each candidate is a statement that may span several lines. To identify the specific defect lines, the validator uses this ranking to choose which complete files to read and in what order. Based on the defect description, fault hypotheses, and complete source, the LLM selects the code lines most likely to contain the root cause. If the source exceeds the input limit, the validator splits it into consecutive parts without omission and checks every part. It removes results with missing files, out-of-range lines, or duplicate lines. When validation requires several batches, the validator restricts reranking to the valid lines returned by these batches. The validator outputs the highly ranked code lines. It ranks files by their first appearance in the line ranking. It maps each line to its containing function and ranks functions in the same way.


\section{EVALUATION}\label{sec:EVALUATION}
In this section, we evaluate LIDL to answer the following research questions (RQs).
\begin{itemize}
    \item \textbf{RQ1:} How effective is LIDL in locating LLM integration defects compared to baselines?
    \item \textbf{RQ2:} How efficient is LIDL in terms of cost?
    \item \textbf{RQ3:} How do different components of LIDL contribute to its effectiveness?

    \item \textbf{RQ4:} How do localization results affect repair effectiveness?

\end{itemize}

\subsection{Experiment Setup}\label{subsec:Experiment Setup}
\begin{figure}[t]
    \centering
    \includegraphics[width=0.36\textwidth]{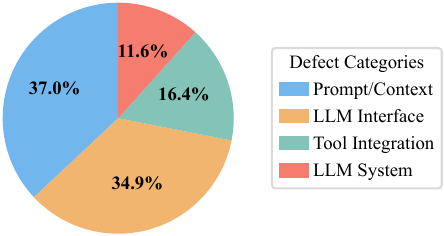}
    \caption{Distribution of the dataset across defect categories.}
    \label{fig:01defect_category_counts}
\end{figure}

\textbf{Dataset.} The dataset contains 146 instances after cleaning from two sources: Hydrangea~\cite{2025ShaoLLM} (888 original defects from 105 GitHub applications) and AgentIssue-Bench~\cite{25RahardjaCan} (50 original defects from 16 agent systems). These include 128 instances from Hydrangea and 18 from AgentIssue-Bench. Fig.~\ref{fig:01defect_category_counts} shows the final distribution of datasets across the four categories.

For data cleaning, we remove instances with (1) missing repository versions on GitHub and (2) incomplete information, such as unclear defect locations. For classification, two annotators independently label all defects into four categories: Prompt and Context Management, LLM Interface Management, Tool Integration Management, and LLM System Management. We exclude instances labeled ``other'' by either annotator and instances with category disagreements. We compute Cohen's kappa~\cite{1960Cohencoefficient} on initial labels and achieve 0.9351, indicating almost perfect agreement.
AgentIssue-Bench provides reference patches. Hydrangea does not, so we construct patches and design defect and regression tests to validate them. A patch passes validation when the defect tests fail before the patch and pass afterward, and the regression tests pass in both versions. We obtain ground truth from the patches of both sources in the same way. The files and lines changed by the patches are the file and line ground truth~\cite{Seddik2026ARISE}. Tree-sitter maps each changed line to its containing function to obtain function ground truth.

\begin{table}[t]
\centering
\caption{Comparison of different methods.}
\label{tab:approach_comparison}
\setlength{\tabcolsep}{0.0001\textwidth}
\resizebox{\linewidth}{!}{%
\begin{tabular}{>{\centering\arraybackslash}m{0.13\textwidth}>{\centering\arraybackslash}m{0.08\textwidth}>{\centering\arraybackslash}m{0.065\textwidth}>{\centering\arraybackslash}m{0.1\textwidth}>{\centering\arraybackslash}m{0.07\textwidth}}
\toprule
Method & Repo Structure & Repo Graph & Domain-Specific Semantic Analysis & Multi Stage \\ \midrule
SWE-agent~\cite{2025YangSWE} & \textcolor{blue}{\ding{52}} & \textcolor{red}{\ding{56}} & \textcolor{red}{\ding{56}} & \textcolor{red}{\ding{56}} \\
Agentless~\cite{2024XiaAgentless} & \textcolor{blue}{\ding{52}} & \textcolor{red}{\ding{56}} & \textcolor{red}{\ding{56}} & \textcolor{blue}{\ding{52}} \\
AutoCodeRover~\cite{2024ZhangAutoCodeRover} & \textcolor{blue}{\ding{52}} & \textcolor{red}{\ding{56}} & \textcolor{red}{\ding{56}} & \textcolor{blue}{\ding{52}} \\
SWE-agent$^*$~\cite{2024OuyangRepoGraph} & \textcolor{blue}{\ding{52}} & \textcolor{blue}{\ding{52}} & \textcolor{red}{\ding{56}} & \textcolor{red}{\ding{56}} \\
Agentless$^*$~\cite{2024OuyangRepoGraph} & \textcolor{blue}{\ding{52}} & \textcolor{blue}{\ding{52}} & \textcolor{red}{\ding{56}} & \textcolor{blue}{\ding{52}} \\
LocAgent~\cite{Chen2025LocAgent} & \textcolor{blue}{\ding{52}} & \textcolor{blue}{\ding{52}} & \textcolor{red}{\ding{56}} & \textcolor{red}{\ding{56}} \\
Codex~\cite{2026OpenAICodex} & \textcolor{blue}{\ding{52}} & \textcolor{red}{\ding{56}} & \textcolor{red}{\ding{56}} & \textcolor{red}{\ding{56}} \\
\midrule
LIDL & \textcolor{blue}{\ding{52}} & \textcolor{blue}{\ding{52}} & \textcolor{blue}{\ding{52}} & \textcolor{blue}{\ding{52}} \\
\bottomrule
\end{tabular}%
}
\end{table}

\vspace{0.1cm}
\noindent
\textbf{Baselines.} We compare LIDL against seven defect localization approaches (\S\ref{sec:PRELIMINARIES}). Table~\ref{tab:approach_comparison} shows their characteristics. Here, repo structure refers to the use of directory, file, class, and function information. Repo graph refers to the use of a graph of code entities and their relations. Domain-specific semantic analysis refers to analyzing code to identify its functional role in LLM workflows. In our method, an LLM analyzes file contents and adds LLM-related annotations to files (\S\ref{sec:semantic-annotation}). Multi Stage indicates that a method uses predefined localization stages in a fixed order. 
\begin{itemize}[leftmargin=*]
    \item \textbf{SWE-agent}~\cite{2025YangSWE} uses an LLM agent to explore codebases and locate defect sources through a custom interface with actions for search, file editing, and context management.
    \item \textbf{Agentless}~\cite{2024XiaAgentless} locates defects through hierarchical localization without agent tools, offering a simple and cost-effective approach.
    \item \textbf{AutoCodeRover}~\cite{2024ZhangAutoCodeRover} provides the LLM agent with code search APIs to find code context and locate defects, supporting class and function-level searching.
    \item \textbf{RepoGraph-enhanced approaches (SWE-agent$^*$ and Agentless$^*$)}~\cite{2024OuyangRepoGraph} add RepoGraph for context.
    \item \textbf{LocAgent}~\cite{Chen2025LocAgent} uses an LLM agent to search a repository graph for code relevant to the defect description.
    \item \textbf{Codex}~\cite{2026OpenAICodex} is a general-purpose coding agent used here for defect localization.
\end{itemize}

We provide all methods with the same defect descriptions and defective repository versions while retaining their search mechanisms. Each method, including ours, returns a ranked list of up to 10 line locations. We use a unified evaluation script to normalize file paths and remove duplicate locations. The script then ranks the files and functions containing the predicted lines by their first appearance in the list.

\vspace{0.1cm}
\noindent
\textbf{Evaluation Metrics.} We evaluate localization at the file, function, and line levels. We use Top-$k$ ($k$=1, 3, 5), Mean Average Precision (MAP), Mean Reciprocal Rank (MRR), Average Cost (\$), and Average Input/Output Tokens (In and Out) to evaluate performance~\cite{22ChenPathidea, 25NiuWhen, 2024XiaAgentless, 2025ChangBridging}.

Top-$k$ measures the percentage of instances with at least one correct location in the top $k$ predictions. MAP computes the mean of Average Precision (AP) across all instances, where AP considers the ranks of all correct locations. MRR computes the mean of the reciprocal rank of the first correct prediction. Average Input Tokens and Average Output Tokens measure the average tokens consumed per instance. Cost is the average price of the LLM calls per instance, computed from the token counts and their prices~\cite{OpenRouter2026Pricing}.
\begin{equation}
    \text{Top-}k = \frac{1}{N}\sum_{i=1}^{N}\mathbf{1}(G^i \cap R^i_{1:k} \neq \emptyset),
\end{equation}
\begin{equation}
    \text{MAP} = \frac{1}{N}\sum_{i=1}^{N}\text{AP}^i,~\text{AP}^i = \frac{1}{|G^i|}\sum_{j=1}^{|R^i|}P^i(j) \cdot \mathbf{1}(r_j^i \in G^i),
\end{equation}
\begin{equation}
    \text{MRR} = \frac{1}{N}\sum_{i=1}^{N}\frac{1}{\text{rank}^i},
\end{equation}
where $N$ is instance count, $G^i$ is the ground-truth set for instance $i$, $R^i = [r_1^i, r_2^i, ...]$ is the ranked prediction list for instance $i$, $R^i_{1:k}$ denotes the top $k$ predictions, $P^i(j) = \frac{|G^i \cap R^i_{1:j}|}{j}$ is precision at rank $j$, $\mathbf{1}(\cdot)$ is the indicator function, and $\text{rank}^i = \min\{j: r_j^i \in G^i\}$ is the position of the first correct prediction. If no correct prediction is found, $\frac{1}{\text{rank}^i} = 0$.

\begin{table*}[t]
\centering
\caption{Effectiveness and efficiency of defect localization methods across backbone LLMs. Top-$k$ is the fraction of instances with at least one correct location in the top $k$ predictions. MAP is Mean Average Precision. MRR is Mean Reciprocal Rank. \$ is average USD per instance. In and Out show average input and output tokens (in thousands) per instance. Best results per model are in \textbf{bold}.}
\label{tab:Performance_comparison}
\setlength{\tabcolsep}{2pt}
\resizebox{\linewidth}{!}{%
\begin{tabular}{ll*{5}{c}c*{5}{c}c*{5}{c}cr@{\hspace{1pt}}r@{\hspace{2pt}}r}
\toprule
\multirow{2}{*}{\textbf{Model}} & \multirow{2}{*}{\textbf{Method}} & \multicolumn{5}{c}{\textbf{File}} & & \multicolumn{5}{c}{\textbf{Function}} & & \multicolumn{5}{c}{\textbf{Line}} & & \multicolumn{3}{c}{\textbf{Cost}} \\
\cline{3-7}\cline{9-13}\cline{15-19}\cline{21-23}
\noalign{\vskip 2pt}
 & & \textbf{Top-1} & \textbf{Top-3} & \textbf{Top-5} & \textbf{MAP} & \textbf{MRR} & & \textbf{Top-1} & \textbf{Top-3} & \textbf{Top-5} & \textbf{MAP} & \textbf{MRR} & & \textbf{Top-1} & \textbf{Top-3} & \textbf{Top-5} & \textbf{MAP} & \textbf{MRR} & & \textbf{In} & \textbf{Out} & \textbf{\$} \\
\midrule
\multirow{8}{*}{Gemma-4-31B} & SWE-agent & 0.5000 & 0.5548 & 0.5616 & 0.4488 & 0.5299 & & 0.3493 & 0.4041 & 0.4110 & 0.2971 & 0.3788 & & 0.1507 & 0.1644 & 0.1644 & 0.0675 & 0.1575 & & 31.69 & 1.04 & 0.002 \\
 & Agentless & 0.4932 & 0.5822 & 0.5822 & 0.4702 & 0.5365 & & 0.3767 & 0.4384 & 0.4452 & 0.3493 & 0.4092 & & 0.2534 & 0.2877 & 0.3014 & 0.1529 & 0.2725 & & 13.39 & 0.34 & 0.001 \\
 & AutoCodeRover & 0.3425 & 0.3699 & 0.3699 & 0.3213 & 0.3562 & & 0.3082 & 0.3219 & 0.3219 & 0.2725 & 0.3151 & & 0.1438 & 0.1644 & 0.1644 & 0.0775 & 0.1541 & & 94.40 & 5.80 & 0.005 \\
 & SWE-agent$^*$ & 0.2945 & 0.3082 & 0.3082 & 0.2660 & 0.3014 & & 0.2260 & 0.2397 & 0.2397 & 0.1924 & 0.2329 & & 0.0548 & 0.0616 & 0.0616 & 0.0176 & 0.0582 & & 42.69 & 1.36 & 0.002 \\
 & Agentless$^*$ & 0.4795 & 0.5479 & 0.5479 & 0.4345 & 0.5126 & & 0.3699 & 0.4041 & 0.4110 & 0.3017 & 0.3864 & & 0.1986 & 0.2877 & 0.3014 & 0.1189 & 0.2428 & & 18.42 & 0.35 & 0.001 \\
 & LocAgent & 0.4521 & 0.5205 & 0.5205 & 0.4358 & 0.4852 & & 0.3767 & 0.4247 & 0.4247 & 0.3416 & 0.4007 & & 0.1986 & 0.2329 & 0.2329 & 0.1042 & 0.2146 & & 580.82 & 15.69 & 0.028 \\
 & Codex & 0.5479 & 0.6438 & 0.6507 & 0.5094 & 0.5982 & & 0.3973 & 0.4658 & 0.4726 & 0.3566 & 0.4323 & & 0.1096 & 0.1301 & 0.1301 & 0.0608 & 0.1199 & & 315.47 & 1.15 & 0.014 \\
 & LIDL & \textbf{0.5616} & \textbf{0.6986} & \textbf{0.7397} & \textbf{0.5818} & \textbf{0.6366} & & \textbf{0.4384} & \textbf{0.5890} & \textbf{0.6301} & \textbf{0.4557} & \textbf{0.5116} & & \textbf{0.3014} & \textbf{0.4041} & \textbf{0.4452} & \textbf{0.2294} & \textbf{0.3597} & & 97.58 & 1.55 & 0.004 \\
\midrule
\multirow{8}{*}{GLM-5.2} & SWE-agent & 0.5000 & 0.6507 & 0.6644 & 0.5210 & 0.5742 & & 0.4041 & 0.5342 & 0.5685 & 0.4050 & 0.4739 & & 0.1849 & 0.2671 & 0.2877 & 0.1282 & 0.2282 & & 101.52 & 2.71 & 0.154 \\
 & Agentless & 0.5000 & 0.5753 & 0.5822 & 0.4827 & 0.5360 & & 0.3836 & 0.4384 & 0.4521 & 0.3525 & 0.4118 & & 0.2192 & 0.3288 & 0.3356 & 0.1737 & 0.2709 & & 13.48 & 0.39 & 0.021 \\
 & AutoCodeRover & 0.3425 & 0.3973 & 0.4041 & 0.3522 & 0.3704 & & 0.3151 & 0.3630 & 0.3699 & 0.2992 & 0.3385 & & 0.1781 & 0.1918 & 0.1986 & 0.0862 & 0.1855 & & 145.04 & 6.06 & 0.230 \\
 & SWE-agent$^*$ & 0.3973 & 0.5342 & 0.5411 & 0.4216 & 0.4629 & & 0.3288 & 0.4452 & 0.4863 & 0.3371 & 0.3889 & & 0.1918 & 0.2671 & 0.2877 & 0.1287 & 0.2305 & & 162.71 & 3.64 & 0.244 \\
 & Agentless$^*$ & 0.5000 & 0.5753 & 0.5753 & 0.4622 & 0.5342 & & 0.3836 & 0.4384 & 0.4384 & 0.3378 & 0.4053 & & 0.1849 & 0.2808 & 0.2945 & 0.1471 & 0.2346 & & 22.43 & 0.41 & 0.033 \\
 & LocAgent & 0.5616 & 0.7397 & \textbf{0.7603} & 0.5952 & 0.6506 & & 0.4178 & 0.6164 & 0.6781 & 0.4557 & 0.5229 & & 0.2123 & 0.3356 & 0.4384 & 0.1914 & 0.3006 & & 641.79 & 2.37 & 0.909 \\
 & Codex & 0.4795 & 0.7123 & 0.7466 & 0.5223 & 0.5919 & & 0.3904 & 0.5479 & 0.6575 & 0.4036 & 0.4893 & & 0.2671 & 0.3836 & 0.4178 & 0.1886 & 0.3322 & & 1124.67 & 3.08 & 1.588 \\
 & LIDL & \textbf{0.5890} & \textbf{0.7603} & \textbf{0.7603} & \textbf{0.6006} & \textbf{0.6701} & & \textbf{0.4795} & \textbf{0.6849} & \textbf{0.7192} & \textbf{0.4851} & \textbf{0.5733} & & \textbf{0.3151} & \textbf{0.4521} & \textbf{0.5205} & \textbf{0.2639} & \textbf{0.4032} & & 76.79 & 1.33 & 0.113 \\
\midrule
\multirow{8}{*}{\shortstack[l]{DeepSeek-\\V4-Flash}} & SWE-agent & 0.4041 & 0.5274 & 0.5479 & 0.4167 & 0.4675 & & 0.3493 & 0.4452 & 0.4795 & 0.3350 & 0.4021 & & 0.1986 & 0.2397 & 0.2534 & 0.1108 & 0.2186 & & 163.85 & 4.57 & 0.078 \\
 & Agentless & 0.4452 & 0.5753 & 0.6096 & 0.4470 & 0.5132 & & 0.3151 & 0.4178 & 0.4589 & 0.3198 & 0.3747 & & 0.1986 & 0.2740 & 0.2945 & 0.1605 & 0.2390 & & 15.77 & 0.45 & 0.008 \\
 & AutoCodeRover & 0.4110 & 0.5342 & 0.5411 & 0.4294 & 0.4706 & & 0.3425 & 0.4452 & 0.4658 & 0.3413 & 0.3926 & & 0.1575 & 0.2055 & 0.2123 & 0.0844 & 0.1798 & & 135.63 & 7.98 & 0.070 \\
 & SWE-agent$^*$ & 0.2877 & 0.3699 & 0.3699 & 0.2931 & 0.3253 & & 0.2260 & 0.3151 & 0.3219 & 0.2303 & 0.2666 & & 0.1233 & 0.1575 & 0.1781 & 0.0739 & 0.1450 & & 146.90 & 3.59 & 0.069 \\
 & Agentless$^*$ & 0.3904 & 0.5068 & 0.5205 & 0.3806 & 0.4498 & & 0.2877 & 0.3836 & 0.4178 & 0.2870 & 0.3365 & & 0.1644 & 0.2260 & 0.2466 & 0.1125 & 0.1995 & & 27.21 & 0.49 & 0.013 \\
 & LocAgent & 0.4726 & 0.6301 & 0.6575 & 0.5043 & 0.5491 & & 0.4110 & 0.5342 & 0.6027 & 0.4175 & 0.4869 & & 0.2123 & 0.3493 & 0.3836 & 0.1480 & 0.2857 & & 801.73 & 14.66 & 0.372 \\
 & Codex & 0.4589 & 0.5753 & 0.5753 & 0.4585 & 0.5171 & & 0.3562 & 0.5068 & 0.5137 & 0.3653 & 0.4272 & & 0.2329 & 0.3219 & 0.3630 & 0.1620 & 0.2858 & & 342.80 & 3.59 & 0.156 \\
 & LIDL & \textbf{0.5411} & \textbf{0.6918} & \textbf{0.7055} & \textbf{0.5418} & \textbf{0.6130} & & \textbf{0.4315} & \textbf{0.5822} & \textbf{0.6301} & \textbf{0.4302} & \textbf{0.5067} & & \textbf{0.2808} & \textbf{0.3904} & \textbf{0.4521} & \textbf{0.2288} & \textbf{0.3457} & & 148.14 & 2.12 & 0.068 \\
\midrule
\multirow{8}{*}{GPT-5.6-Terra} & SWE-agent & 0.3562 & 0.4452 & 0.4521 & 0.3609 & 0.3990 & & 0.2877 & 0.3767 & 0.3836 & 0.2779 & 0.3279 & & 0.1301 & 0.1575 & 0.1575 & 0.0727 & 0.1448 & & 56.78 & 0.58 & 0.121 \\
 & Agentless & 0.5205 & 0.6438 & 0.6507 & 0.5245 & 0.5816 & & 0.4110 & 0.5274 & 0.5479 & 0.4045 & 0.4694 & & 0.2397 & 0.3699 & 0.3973 & 0.1625 & 0.3030 & & 16.26 & 0.31 & 0.036 \\
 & AutoCodeRover & 0.5411 & 0.6096 & 0.6233 & 0.5198 & 0.5784 & & 0.4384 & 0.5137 & 0.5342 & 0.3961 & 0.4797 & & 0.1918 & 0.1986 & 0.1986 & 0.0853 & 0.1962 & & 110.72 & 2.83 & 0.255 \\
 & SWE-agent$^*$ & 0.3836 & 0.5274 & 0.5342 & 0.4036 & 0.4526 & & 0.3356 & 0.4932 & 0.5068 & 0.3370 & 0.4087 & & 0.1986 & 0.2808 & 0.3014 & 0.1038 & 0.2380 & & 63.78 & 0.74 & 0.137 \\
 & Agentless$^*$ & 0.4726 & 0.5959 & 0.6027 & 0.4613 & 0.5337 & & 0.3288 & 0.4658 & 0.5000 & 0.3317 & 0.3975 & & 0.1301 & 0.2192 & 0.2466 & 0.0999 & 0.1837 & & 24.08 & 0.35 & 0.052 \\
 & LocAgent & 0.4795 & 0.6370 & 0.6575 & 0.4967 & 0.5611 & & 0.3699 & 0.4932 & 0.5548 & 0.3737 & 0.4464 & & 0.1575 & 0.2192 & 0.3014 & 0.1133 & 0.2082 & & 755.56 & 2.49 & 1.541 \\
 & Codex & 0.4726 & 0.6301 & 0.6438 & 0.4815 & 0.5484 & & 0.4452 & 0.5685 & 0.5890 & 0.4204 & 0.5064 & & 0.2671 & 0.4110 & 0.4178 & 0.1626 & 0.3358 & & 148.73 & 1.70 & 0.318 \\
 & LIDL & \textbf{0.5890} & \textbf{0.7671} & \textbf{0.7808} & \textbf{0.6055} & \textbf{0.6758} & & \textbf{0.4932} & \textbf{0.6918} & \textbf{0.7260} & \textbf{0.4967} & \textbf{0.5901} & & \textbf{0.3151} & \textbf{0.5000} & \textbf{0.5479} & \textbf{0.2436} & \textbf{0.4135} & & 46.90 & 1.06 & 0.107 \\
\bottomrule
\end{tabular}%
}
\end{table*}

\vspace{0.1cm}
\noindent
\textbf{Configurations.} We evaluate all methods with Gemma-4-31B~\cite{Abd2026Gemma4}, GLM-5.2~\cite{GLM2026GLM5}, DeepSeek-V4-Flash~\cite{DeepSeek2026V4}, and GPT-5.6-Terra~\cite{OpenAI2026GPT56Terra}. Methods that require embeddings use BGE-M3~\cite{2024ChenM3}. We set $temperature=0.0$ for all methods and LLMs that support it. All experiments run with Ubuntu 24.04, 64-core Intel Xeon Gold 6326 CPU, 128GB RAM, and 6 NVIDIA A40 48GB GPUs. We implement all methods using Python 3.11.15.

\vspace{0.1cm}
\noindent
\textbf{Parameter Settings.} For LLM-related annotation, we use $k_s=5$ seed files and add up to $k_e=5$ files through graph expansion. These budgets balance files matched by regular expressions with additional files recalled by graph expansion under the LLM's input limit. For file rank fusion, we use weighted reciprocal rank fusion with the smoothing constant $k=60$ to control score differences between file ranks~\cite{Liu25Exp4Fuse}. We set the base weights $w_r$ to $2$, $1$, $2$, and $2$ for symptom-based, annotation-based, text-based, and graph-based retrieval. To choose them, we randomly sample 30 instances from the evaluation set in proportion to defect categories and test 16 weight combinations with values from $\{1,2\}$. The selected combination gives the best file Top-5 accuracy.
 For baselines, we use their original default parameters.

\subsection{RQ1: Effectiveness in Defect Localization}
Table~\ref{tab:Performance_comparison} reports the file, function, and line localization results of all methods on each backbone LLM. Each row gives the results of a method on a backbone LLM. The columns give Top-$k$, MAP, and MRR at each level, followed by tokens and cost, which we analyze in RQ2. We first compare localization performance across backbone LLMs.
We then use GPT-5.6-Terra for overall comparison, cross-category performance analysis, and overlap analysis.
We analyze Top-3 because developers may inspect several candidate locations rather than only the first prediction.

\vspace{0.1cm}
\noindent
\textbf{Model Comparison.} 
On every backbone LLM, our file, function, and line Top-3 are higher than those of all baselines.
Some other metrics show smaller differences. With GLM-5.2, our file MAP of 0.6006 is close to LocAgent's 0.5952, and both methods achieve file Top-5 of 0.7603.
To summarize performance across backbone LLMs, we report each method's line Top-3 as mean $\pm$ standard deviation.
We achieve the highest average at $0.4366 \pm 0.0432$.
Agentless and Codex follow at $0.3151 \pm 0.0375$ and $0.3116 \pm 0.1097$.
LocAgent averages $0.2843 \pm 0.0586$, and Agentless$^*$ averages $0.2534 \pm 0.0310$.
SWE-agent averages $0.2072 \pm 0.0473$, and SWE-agent$^*$ averages $0.1918 \pm 0.0891$.
AutoCodeRover has the lowest average at $0.1901 \pm 0.0156$.
The backbone LLM with the highest line Top-3 differs by method.
DeepSeek-V4-Flash gives the highest score for AutoCodeRover and LocAgent.
GLM-5.2 gives the highest score for SWE-agent, and Gemma-4-31B gives the highest score for Agentless$^*$.
GPT-5.6-Terra gives the highest score for our method, Agentless, SWE-agent$^*$, and Codex.
GPT-5.6-Terra also gives the highest function Top-3 for most methods, so we use it for the following analysis.

\vspace{0.1cm}
\noindent
\textbf{Overall Performance Comparison.} On GPT-5.6-Terra, we outperform the compared baselines on all localization metrics reported in Table~\ref{tab:Performance_comparison}. Our file Top-3 is 0.7671, a relative improvement of 19.15\% over Agentless (0.6438), which has the highest baseline score on this metric. Our function and line Top-3 scores are 0.6918 and 0.5. They exceed the corresponding highest baseline scores from Codex (0.5685 and 0.411) by 21.69\% and 21.65\%. Our file, function, and line MAP scores of 0.6055, 0.4967, and 0.2436 also exceed the corresponding highest baseline values of 0.5245, 0.4204, and 0.1626. We analyze which components contribute to this gain in RQ3 (\S\ref{subsec:rq3}).

\begin{figure}[t]
    \centering
    \includegraphics[width=0.489\textwidth]{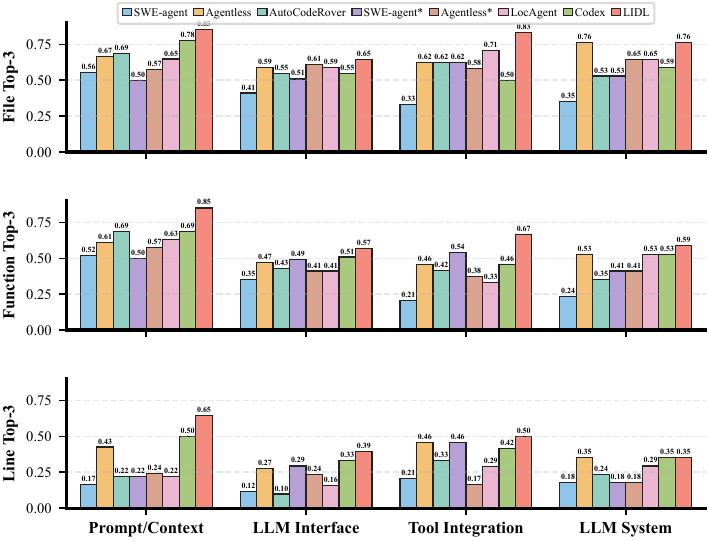}
    \caption{Effectiveness comparison of methods across different defect categories. Top-3 at the file, function, and line levels. Backbone LLM: GPT-5.6-Terra.}
    \label{fig:8_defect_category_methods_comparison_combined}
    \vspace{-4mm}
\end{figure}

\vspace{0.1cm}
\noindent
\textbf{Performance on Different Defect Categories.} Fig.~\ref{fig:8_defect_category_methods_comparison_combined} shows Top-3 at the file, function, and line levels for each defect category. We achieve higher function Top-3 across defect categories. In Prompt and Context Management, our file, function, and line Top-3 scores all exceed those of the baselines. Our line Top-3 is 0.6481, a relative improvement of 29.62\% over the highest baseline score of 0.5. In LLM Interface Management, we also achieve higher Top-3 at every level, with line Top-3 of 0.3922 exceeding the highest baseline score of 0.3333 by 17.67\%. In Tool Integration Management, Top-3 is again higher at every level, with line Top-3 of 0.5 exceeding the highest baseline score of 0.4583 by 9.1\%. 
In LLM System Management, our file and line Top-3 match the highest baseline scores, while our function Top-3 of 0.5882 exceeds the highest baseline score of 0.5294.

\begin{figure}[t]
    \centering
    \includegraphics[width=0.489\textwidth]{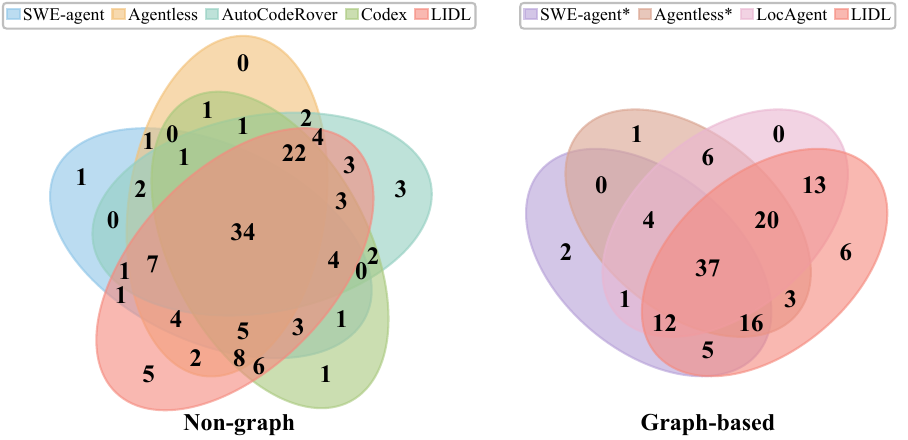}
    \caption{Overlap of localized defects among methods at file-level Top-3. Backbone LLM: GPT-5.6-Terra.}
    \label{fig:venn_method_overlap}
    \vspace{-3mm}
\end{figure}

\begin{figure}[t]
    \centering
    \includegraphics[width=0.489\textwidth]{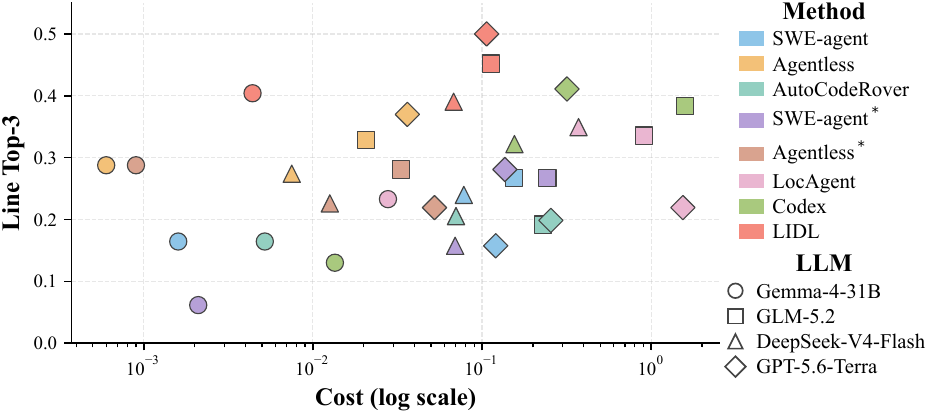}
    \caption{Average cost per instance (USD) and line-level defect localization accuracy (Top-3) across methods and backbone LLMs.}
    \label{fig:cost_accuracy_scatter}
    \vspace{-4mm}
\end{figure}

\vspace{0.1cm}
\noindent
\textbf{Overlap Analysis.} We compare the defects located by each method to identify common and unique results. A defect is counted as located when the top 3 predicted files include at least one correct file. Fig.~\ref{fig:venn_method_overlap} compares our results separately with non-graph baselines (SWE-agent, Agentless, AutoCodeRover, and Codex) and graph-based baselines (SWE-agent$^*$, Agentless$^*$, and LocAgent). We locate 112 defects, accounting for 76.71\% of all instances. Of these, 34 defects in the non-graph group and 37 in the graph-based group are located by every method in the same group. Of the 112 defects, 5 are missed by every baseline in the non-graph group, and 6 are missed by every baseline in the graph-based group. The baselines also locate defects that we miss. For example, AutoCodeRover uniquely locates 3 defects in the non-graph group.

\begin{tcolorbox}[boxsep=1pt,left=2pt,right=2pt,top=3pt,bottom=2pt,width=\linewidth,colback=white!95!black,boxrule=1pt, colbacktitle=white!30!black,toptitle=2pt,bottomtitle=1pt,opacitybacktitle=0.4]
\textbf{Answer to RQ1.} We localize defects more accurately than the compared baselines at the file, function, and line levels across the evaluated backbone LLMs. On GPT-5.6-Terra, our Top-3 exceeds the best baseline by 19.15\% at the file level, 21.69\% at the function level, and 21.65\% at the line level. We also achieve higher function Top-3 across defect categories.   
 In addition, we identify defects missed by all non-graph baselines and defects missed by all graph-based baselines.
\end{tcolorbox}

\subsection{RQ2: Efficiency Analysis}

We analyze efficiency through token consumption and the corresponding LLM call costs.
Table~\ref{tab:Performance_comparison} lists the average input tokens, output tokens, and cost of each method per instance.
Fig.~\ref{fig:cost_accuracy_scatter} plots the cost and line Top-3 in the table.
Cost combines input and output tokens with their prices, so we use it as the main measure.
Across the evaluated models, our cost is lower than that of AutoCodeRover, LocAgent, and Codex, higher than that of Agentless and Agentless$^*$, and lower than that of SWE-agent and SWE-agent$^*$ on most models. Taking GPT-5.6-Terra as an example, we use an average of 46.9k input tokens and 1.06k output tokens per instance. Compared with AutoCodeRover (2.83k), LocAgent (2.49k), and Codex (1.7k), we reduce output tokens by 62.58\%, 57.51\%, and 37.65\%. SWE-agent and SWE-agent$^*$ also use fewer output tokens, at 0.58k and 0.74k, but use more input tokens, at 56.78k and 63.78k. Our average cost is \$0.107 per instance, which is 58.3\%, 93.09\%, and 66.5\% lower than that of AutoCodeRover, LocAgent, and Codex. It is also lower than that of SWE-agent (\$0.121) and SWE-agent$^*$ (\$0.137), despite their lower output token counts. Agentless and Agentless$^*$ cost less, at \$0.036 and \$0.052, but their file Top-3 scores of 0.6438 and 0.5959 are lower than our score of 0.7671. Thus, we achieve higher localization accuracy at a lower cost than most baselines.

\begin{tcolorbox}[boxsep=1pt,left=2pt,right=2pt,top=3pt,bottom=2pt,width=\linewidth,colback=white!95!black,boxrule=1pt, colbacktitle=white!30!black,toptitle=2pt,bottomtitle=1pt,opacitybacktitle=0.4]
\textbf{Answer to RQ2.} Our cost is lower than that of AutoCodeRover, LocAgent, and Codex on all evaluated models, and lower than that of SWE-agent and SWE-agent$^*$ on most models. On GPT-5.6-Terra, our cost of \$0.107 per instance is 58.3\%, 93.09\%, and 66.5\% lower than that of AutoCodeRover, LocAgent, and Codex. Agentless and Agentless$^*$ cost less, but we achieve higher localization accuracy than both.
\end{tcolorbox}

\begin{table}[t]
\centering
\caption{Ablation study of retrieval mechanisms and validation. $R_e$: direct extraction. $R_t$: text-based retrieval. $R_i$: symptom-based retrieval. $R_r$: annotation-based retrieval. $R_g$: graph-based retrieval. $\mathcal{V}$: validator. w/o $(\mathcal{V} \,\&\, R)$ removes the validator and one retrieval source. w/o non-code files removes predicted locations in non-code files from the final output. Backbone LLM: GPT-5.6-Terra.}
\label{tab:ablation}
\setlength{\tabcolsep}{2.5pt}
\resizebox{\linewidth}{!}{%
\begin{tabular}{l*{3}{c}c*{3}{c}c*{3}{c}}
\toprule
\multicolumn{1}{c}{\multirow{2}{*}{Method}} & \multicolumn{3}{c}{File} & & \multicolumn{3}{c}{Function} & & \multicolumn{3}{c}{Line} \\
\cline{2-4}\cline{6-8}\cline{10-12}
\noalign{\vskip 2pt}
 & Top-1 & Top-3 & MAP & & Top-1 & Top-3 & MAP & & Top-1 & Top-3 & MAP \\
\midrule
w/o $(\mathcal{V} \,\&\, R_e)$ & 0.2329 & 0.3356 & 0.2393 & & 0.1233 & 0.1849 & 0.1114 & & 0.0205 & 0.0342 & 0.0134 \\
w/o $(\mathcal{V} \,\&\, R_t)$ & 0.2877 & 0.4452 & 0.3018 & & 0.1918 & 0.2534 & 0.1683 & & 0.0274 & 0.0411 & 0.0121 \\
w/o $(\mathcal{V} \,\&\, R_i)$ & 0.2123 & 0.3082 & 0.2163 & & 0.1027 & 0.1301 & 0.0909 & & 0.0068 & 0.0205 & 0.0078 \\
w/o $(\mathcal{V} \,\&\, R_r)$ & 0.2466 & 0.3288 & 0.2416 & & 0.1370 & 0.1781 & 0.1257 & & 0.0205 & 0.0411 & 0.0199 \\
w/o $(\mathcal{V} \,\&\, R_g)$ & 0.2603 & 0.4247 & 0.2882 & & 0.1575 & 0.2192 & 0.1459 & & 0.0274 & 0.0411 & 0.0118 \\
\midrule
w/o $\mathcal{V}$ & 0.2534 & 0.3562 & 0.2553 & & 0.1507 & 0.2055 & 0.1354 & & 0.0205 & 0.0411 & 0.0200 \\
\midrule
w/o non-code files & 0.5685 & 0.7397 & 0.5896 & & 0.4658 & 0.6644 & 0.4704 & & 0.3082 & 0.4863 & 0.2309 \\
\midrule
LIDL & 0.5890 & 0.7671 & 0.6055 & & 0.4932 & 0.6918 & 0.4967 & & 0.3151 & 0.5000 & 0.2436 \\
\bottomrule
\end{tabular}%
}
\end{table}

\subsection{RQ3: Ablation Study of LIDL Components}
\label{subsec:rq3}
Table~\ref{tab:ablation} presents the ablation results on GPT-5.6-Terra. Without the validator (w/o $\mathcal{V}$), the output is the analyzer's candidate files and candidate code lines in their initial ranking. We compare this result with the full method to assess the validator. To assess each retrieval source, we remove the validator and the source together (w/o $(\mathcal{V} \,\&\, R)$) and compare the result with w/o $\mathcal{V}$. We also remove predicted locations in non-code files from the final output (w/o non-code files) and compare the result with the full method.


\vspace{0.1cm}\noindent\textbf{Effect of the Validator.} Localization results after validation are more accurate than the retrieval results before validation. File, function, and line Top-3 increase from 0.3562, 0.2055, and 0.0411 to 0.7671, 0.6918, and 0.5. The corresponding MAP scores increase from 0.2553, 0.1354, and 0.02 to 0.6055, 0.4967, and 0.2436. These results show that the validator improves both localization accuracy and the ranking of correct locations.

\vspace{0.1cm}\noindent\textbf{Effect of Retrieval Evidence.} Direct extraction, symptom-based retrieval, and annotation-based retrieval evidence each contribute to file and function localization before validation. Removing direct extraction evidence reduces file, function, and line Top-3 to 0.3356, 0.1849, and 0.0342. Removing symptom-based retrieval evidence reduces the corresponding scores to 0.3082, 0.1301, and 0.0205. Removing annotation-based retrieval evidence reduces file and function Top-3 to 0.3288 and 0.1781. Among these variants, removing symptom-based retrieval evidence causes the largest decrease.

Removing text-based retrieval increases file and function Top-3 before validation, and so does removing graph-based retrieval. Without text-based retrieval, file and function Top-3 are 0.4452 and 0.2534. Without graph-based retrieval, they are 0.4247 and 0.2192. These scores exceed 0.3562 and 0.2055 with all retrieval sources retained. Text-based retrieval matches words, and graph-based retrieval follows relations between files. Both explore additional files related to the defect description. These files do not always contain the root cause and can rank above correct files during candidate aggregation. We keep these retrieval methods to supplement candidates that other retrieval methods may miss. The validator then uses code context to rerank the candidates.

\vspace{0.1cm}\noindent\textbf{Effect of Non-code Files.} In the ground truth, 13 of 146 instances (8.9\%) involve non-code files. Among the final predictions, 14 instances include locations in non-code files. Removing these locations lowers file, function, and line Top-3 from 0.7671, 0.6918, and 0.5 to 0.7397, 0.6644, and 0.4863. These results show that non-code files contribute to localization accuracy.

\begin{tcolorbox}[boxsep=1pt,left=2pt,right=2pt,top=3pt,bottom=2pt,width=\linewidth,colback=white!95!black,boxrule=1pt, colbacktitle=white!30!black,toptitle=2pt,bottomtitle=1pt,opacitybacktitle=0.4]
\textbf{Answer to RQ3.} The validator improves both localization accuracy and the ranking of correct locations, raising file, function, and line Top-3 from 0.3562, 0.2055, and 0.0411 to 0.7671, 0.6918, and 0.5. Among the retrieval evidence before validation, direct extraction, symptom-based retrieval, and annotation-based retrieval each improve file and function localization, and symptom-based retrieval contributes the most. Text-based and graph-based retrieval provide additional candidates, which the validator then reranks. Non-code files also improve localization at the file, function, and line levels.

\end{tcolorbox}

\subsection{RQ4: Impact on Repair Effectiveness}
We use the defect description and the files containing the lines predicted by each localization method as input for repair. For all methods, we use the same LLM (GPT-5.6-Terra), prompt template, and budget. We then evaluate the generated patches using the corresponding defect and regression tests. We measure repair success as the fraction of instances whose patches pass all specified tests, denoted as Pass@1. Table~\ref{tab:repair_comparison} reports the results.

\begin{table}[t]
\centering
\caption{Downstream repair using context from different localization methods. Resolved counts instances whose patches pass all tests. Pass@1 (\%) is calculated as Resolved/146 $\times$ 100.}
\label{tab:repair_comparison}
\setlength{\tabcolsep}{2.5pt}
\begin{tabular}{l c c}
\toprule
Method & Pass@1 & Resolved \\
\midrule
SWE-agent & 8.22 & 12 \\
Agentless & 8.90 & 13 \\
AutoCodeRover & 8.90 & 13 \\
SWE-agent$^*$ & 5.48 & 8 \\
Agentless$^*$ & 8.22 & 12 \\
LocAgent & 7.53 & 11 \\
Codex & 6.16 & 9 \\
\midrule
LIDL & \textbf{11.64} & \textbf{17} \\
\bottomrule
\end{tabular}
\end{table}

Our localization results achieve the highest repair success rate. We achieve a Pass@1 of 11.64\%, resolving 17 instances. Agentless and AutoCodeRover are the best-performing baselines, each achieving a Pass@1 of 8.9\% and resolving 13 instances. We resolve 4 more instances than these baselines, a relative improvement of 30.77\% in repair success. These results show that our localization better supports downstream repair.

\begin{tcolorbox}[boxsep=1pt,left=2pt,right=2pt,top=3pt,bottom=2pt,width=\linewidth,colback=white!95!black,boxrule=1pt, colbacktitle=white!30!black,toptitle=2pt,bottomtitle=1pt,opacitybacktitle=0.4]
\textbf{Answer to RQ4.} Under the same repair LLM, prompt template, and budget, our localization results lead to higher repair success than those of the compared baselines. We achieve a Pass@1 of 11.64\% and resolve 17 instances, a relative improvement of 30.77\% over the best-performing baselines.
\end{tcolorbox}

\section{DISCUSSION}\label{sec:DISCUSSION}

\subsection{Implications}
In this section, we discuss how our results help developers, users, and researchers.

\textbf{Symptoms in the defect description are an effective clue for retrieval.} A defect description contains file names and code symbols, and it also describes the abnormal behavior of the program, called symptoms. In the ablation before validation, removing symptom-based retrieval lowers file and function localization accuracy more than removing any other retrieval source (Table~\ref{tab:ablation}). This shows that symptom-based retrieval provides additional evidence for locating defects. Tool developers should therefore use symptom-based retrieval together with direct extraction and annotation-based retrieval to obtain candidate files.

\textbf{The file ranking from retrieval needs validation.} Retrieval ranks candidate files, but the top-ranked files do not always contain the defect. Our validator reads the code context of the candidate files and judges whether correcting each file could remove the symptoms. It uses this judgment to guide the ranking of candidate code lines. After validation, file Top-3 rises from 0.3562 to 0.7671 (Table~\ref{tab:ablation}). Users read code in the order of the ranking, so they read the top-ranked files first. Tool developers should therefore consider validating file rankings after retrieval rather than using retrieval rankings directly.

\textbf{Multi-stage tools should evaluate each stage.} The ablation shows that direct extraction, symptom-based retrieval, and annotation-based retrieval each improve file and function localization before validation, while text-based and graph-based retrieval do not improve file and function Top-3 before validation. Adding validation on top of retrieval raises accuracy further (Table~\ref{tab:ablation}). Tool developers should therefore evaluate each stage and explain what problem each stage solves.

\textbf{Tool choice needs a balance between localization accuracy and cost.} A method with higher accuracy does not always cost less, and the accuracy and cost of each method change with the LLM (Table~\ref{tab:Performance_comparison}). Users should therefore compare accuracy and cost under the LLM they plan to use. For example, with GPT-5.6-Terra, Agentless costs \$0.036 per instance and reaches a file Top-3 of 0.6438, while our method costs \$0.107 and reaches 0.7671.

\textbf{Localization accuracy should be evaluated together with repair results.} We use the localization results of each method for repair under the same setup. Repair with our localization results fixes 17 defects, and repair with the best baseline fixes 13 (Table~\ref{tab:repair_comparison}). The methods rank differently by localization accuracy and by repair success. After the correct location is found, a patch may still fail to fix the defect, and tests need to confirm it. Localization accuracy therefore does not show whether the final repair succeeds. Researchers should report localization accuracy and repair test results together.

\textbf{Defect detection before running and defect localization after failure are complementary.} Some defects in LLM-integrated software are similar to traditional software defects (Section~\ref{subsec:IntegrationDefect}). In Case 2, an empty tool result is written into the message, and the API returns an error. In Case 3, importing a toolkit also imports other toolkits, so unrelated toolkits fail to import when optional dependencies are missing. In Case 4, the dependency declaration lacks tiktoken, and a new environment cannot start. We can detect these defects before running~\cite{Du2026Reducing}, and LLM integration does not prevent this. Case 1 is different. A prompt is missing, and the system raises no error and picks the wrong sub-agent. Detection before running may miss this kind of defect, while defect localization can identify its location after the failure is observed. Developers of LLM-integrated software should therefore keep defect detection before running and add defect localization.

\subsection{Limitations and Future Work}
Our work has four limitations. (1) The localization cost of LIDL varies with model pricing, ranging from \$0.004 to \$0.113 per instance, although it maintains a cost advantage over AutoCodeRover, LocAgent, and Codex across all evaluated models. Future work could use smaller models for initial filtering. (2) Our code knowledge graph relies on static analysis and may miss dynamic call relations, which can affect defect localization. In addition, we use predefined patterns and LLM-extracted keywords to select files for annotation and retrieval. The predefined patterns come from popular LLM frameworks, including LangChain, LlamaIndex, and AutoGen. These patterns may miss relevant files in projects using custom or less common frameworks. For LLM-extracted keywords, validation and prefix filtering may change which files are retrieved and how they are ranked. Future work could expand pattern coverage and evaluate how keyword processing affects localization accuracy. (3) LIDL currently supports Python only. The core design, including the code knowledge graph, candidate aggregation, and counterfactual reasoning, can be applied to other languages, but extending to them requires adapting Tree-sitter queries and pattern libraries. Future work could evaluate LIDL on multi-language benchmarks. (4) We run the batched annotation before localization for each repository version, and we exclude its cost from localization costs. Some baselines optimize inputs during localization through code compression (Agentless) and context trimming (SWE-agent). Splitting oversized files during offline annotation may separate related code and limit the context visible to the model. Future work will explore input processing methods that preserve more relevant code context and evaluate their effects on cost and defect localization.

\section{THREATS TO VALIDITY}\label{sec:THREATS}

\vspace{0.1cm}
\noindent
\textbf{Internal Validity.} (1) Dataset reduction through manual filtering may introduce selection bias. We mitigate this risk by maintaining diversity across defect categories, employing two independent annotators with high agreement (Cohen’s kappa: 0.9351), and removing defects by the criteria in Section~\ref{subsec:Experiment Setup}. (2) Ground truth labels may vary across different fixing strategies. We follow the widely used standard from SWE-bench, where modified or deleted code in patches is labeled as buggy~\cite{2024XiaAgentless, 2025ChangBridging}. For Hydrangea, we construct the patches and validate them with our own defect and regression tests, so its ground truth depends on these patches. (3) Each method runs once per instance, and LLM outputs may vary across runs. (4) The retrieval weights are selected using 30 instances from the evaluation set, which may introduce bias into the reported results.

\vspace{0.1cm}
\noindent
\textbf{External Validity.} (1) Our dataset focuses on Python applications from GitHub repositories, which may not represent industrial codebases with different development practices. (2) Although we evaluate several LLMs, performance may vary with other LLMs. However, the benefits across the evaluated LLMs suggest that the advantages may transfer. (3) The defect distribution in our dataset may differ from real-world distributions. Future work could validate LIDL on industrial datasets. (4) The defects come from public GitHub repositories, so the LLMs may have seen them in training. This may affect the evaluation of the compared methods.

\section{CONCLUSION}\label{sec:CONCLUSION}
In conclusion, this work addresses the challenge of localizing LLM integration defects. These defects exhibit three key characteristics that existing methods cannot handle: defects span heterogeneous components beyond source code, error traces point to invocation layers rather than root causes, and defects involve semantic dependencies that require contextual reasoning. To address these challenges, we present LIDL, a multi-agent framework for localizing LLM integration defects. LIDL operates through three coordinated agents: a code knowledge graph constructor that builds a knowledge graph capturing both program structure and LLM interaction points with semantic annotations, a defect analyzer that fuses complementary evidence sources, and a context-aware validator that applies counterfactual reasoning to distinguish root causes from symptoms. Our evaluation shows that LIDL outperforms existing approaches at the file, function, and line levels, improving Top-3 accuracy by 19.15\%, 21.69\%, and 21.65\%. LIDL also achieves lower costs than most baselines and improves downstream repair success by 30.77\%. LIDL provides a novel solution for locating LLM integration defects and improving the reliability of LLM-integrated software development. 

\renewcommand{\refname}{REFERENCES}
\bibliographystyle{IEEEtran}
\bibliography{references}



\end{document}